\documentclass[a4paper,fleqn,usenatbib]{mnras}

\usepackage[T1]{fontenc}
\usepackage{ae,aecompl}

\usepackage{graphicx}	
\usepackage{amsmath}	
\usepackage{amssymb}	
\usepackage{float}
\usepackage{natbib}
\usepackage{morefloats}
\usepackage{dcolumn}
\usepackage{amsmath}
\usepackage{latexsym}
\usepackage{longtable}
\usepackage{siunitx}
\usepackage{adjustbox}
\usepackage{lipsum} 

\hypersetup{draft}
\begin{document}
\title[Optical studies of two stripped envelope supernovae $-$ SN 2015ap (Type Ib) and SN 2016P (Type Ic)]
{Optical studies of two stripped envelope supernovae $-$ SN 2015ap (Type Ib) and SN 2016P (Type Ic)}
\author[Gangopadhyay Anjasha et al.]
{Anjasha Gangopadhyay\thanks{e-mail :anjasha@aries.res.in, anjashagangopadhyay@gmail.com}$^{1,2}$, Kuntal Misra$^{1}$, D. K. Sahu$^{3}$, Shan-Qin Wang $^{4}$,
\newauthor
Brajesh Kumar$^{3}$, Long Li $^{4}$, G. C. Anupama$^{3}$, Raya Dastidar$^{1,5}$, N. Elias-Rosa$^{6,7}$,
\newauthor
Brijesh Kumar$^{1}$, Mridweeka Singh$^{1,2,12}$, S. B. Pandey$^{1}$, Pankaj Sanwal$^{1,2}$,
\newauthor
Avinash Singh$^{3,8}$, S. Srivastav$^{9}$, L. Tartaglia$^{10}$, L. Tomasella$^{11}$     \\
1. Aryabhatta Research Institute of observational sciencES, Manora Peak, Nainital 263 002 India \\
2. School of Studies in Physics and Astrophysics, Pandit Ravishankar Shukla University, Chattisgarh 492 010, India \\
3. Indian Institute of Astrophysics, Koramangala, Bangalore 560 034, India \\
4. Guangxi Key Laboratory for Relativistic Astrophysics, School of Physical Science and Technology, Guangxi University, Nanning 530004, \\
People's Republic of China \\
5. Department of Physics $\&$ Astrophysics, University of Delhi, Delhi-110 007 \\
6. Institute of Space Sciences (ICE, CSIC), Campus UAB, Carrer de Can Magrans s/n, 08193 Barcelona, Spain \\
7. Institut d'Estudis Espacials de Catalunya (IEEC), c/Gran Capit\'a 2-4, Edif. Nexus 201, 08034 Barcelona, Spain \\
8. Joint Astronomy Programme, Department of Physics, Indian Institute of Science, Bengaluru 560012, India \\
9. Queen's University, University Rd, Belfast BT7 1NN, UK \\
10. Department of Astronomy and The Oskar Klein Centre, AlbaNova University Center,Stockholm University, SE-106 91, Stockholm, Sweden  \\
11. INAF Osservatorio Astronomico di Padova, Vicolo dell'Osservatorio 5, 35122 Padova, Italy \\
12. Korea Astronomy and Space Science Institute, 776 Daedeokdae-ro, Yuseong-gu, Daejeon 34055, Republic of Korea}


\date{Accepted XXX. Received YYY; in original form ZZZ}

\pubyear{2016}

\label{firstpage}
\pagerange{\pageref{firstpage}--\pageref{lastpage}}
\maketitle

\begin{abstract}
We present the photometric and spectroscopic studies of a Type Ib SN 2015ap and a Type Ic SN 2016P. SN 2015ap is one of the bright (M$_{V}$ = $-$18.04 mag) Type Ib while SN 2016P lies at an average value among the Type Ic SNe (M$_{V}$ = $-$17.53 mag). Bolometric light curve modelling of SNe 2015ap and 2016P indicates that both the SNe are powered by $^{56}$Ni + magnetar model with $^{56}$Ni masses of 0.01 M$_{\odot}$ and 0.002 M$_{\odot}$, ejecta masses of 3.75 M$_{\odot}$ and 4.66 M$_{\odot}$, spin period P$_{0}$ of 25.8 ms and 36.5 ms and magnetic field B$_{p}$ of 28.39 $\times$ 10$^{14}$ Gauss and 35.3 $\times$ 10$^{14}$ Gauss respectively. The early spectra of SN 2015ap shows prominent lines of He with a ``W" feature due to Fe complexes while other lines of Mg II, Na I and Si II are present in both SNe 2015ap and 2016P. Nebular phase [O I] profile indicates an asymmetric profile in SN 2015ap. The [O I]/[Ca II] ratio and nebular spectral modelling of SN 2015ap hints towards a progenitor mass between 12 $-$ 20 M$_{\odot}$.
\end{abstract}




\section{Introduction}
Massive stars undergoing catastrophic explosions at the end of their life are coined as core-collapse supernovae \citep [CCSNe;][]{1986BAAS...18.1016W,2003ApJ...591..288H,2009ARA&A..47...63S}. Among the group of CCSNe, lies the stripped envelope (SE-SNe) members called SNe of Type IIb, Ib and Ic which strip off their outer Hydrogen or Helium layer before the explosion \citep{1997ApJ...491..375C}. Stripping occurs either due to radiation driven stellar winds \citep{2008A&ARv..16..209P}, heavy mass-loss \citep{2006ApJ...645L..45S}, or mass-transfer with a companion star \citep{1985ApJ...294L..17W,1992ApJ...391..246P,2010ApJ...725..940Y}. Classically, Type Ib SNe show He features and Type Ic are devoid of He features \citep{1997ARA&A..35..309F}. However, studies by \cite{1995ApJ...448..315W,2012MNRAS.424.2139D,2014MNRAS.438.2924C} suggest that in Type Ic, either Helium is hidden because of improper excitation due to poor mixing or is completely absent. A small fraction of Type Ic SNe \citep[$\sim$ 4 $\%$;][]{2017MNRAS.471.4381S} show broad absorption lines in their spectra around maximum light with expansion velocities between 15,000 km s$^{-1}$ to 30,000 km s$^{-1}$. These are known as broad-lined Type Ic SNe (BL-Ic). Some of the BL-Ic SNe are linked with gamma ray bursts (GRBs) and X-ray flashes (XRFs) (for example SN 1998bw/GRB 980425; \cite{1999A&AS..138..465G}, SN 2006aj/XRF 060218; \cite{2006ApJ...645L..21M}).

A number of studies have been made by several authors to investigate the photometric \citep{2006AJ....131.2233R,2011ApJ...741...97D,2013MNRAS.434.1098C,2014ApJS..213...19B,2014ApJ...787..157P,2015A&A...574A..60T,2016MNRAS.457..328L,2016MNRAS.458.2973P} and spectroscopic properties of SE-SNe \citep{2001PASP..113.1155M,2016ApJ...832..108M,2019MNRAS.485.1559P}. These studies have shown a wide diversity in terms of both light curve shapes and spectral features in SE-SNe. The typical rise time of SE-SNe is found to be 10$-$20 days \citep{2019MNRAS.485.1559P} with weighted peak absolute magnitude of M$_{V}$ = $-$18.03 $\pm$ 0.06 mag \citep{2006AJ....131.2233R}. \cite{2016MNRAS.457..328L} found a considerable spread of $\sim$ 3 mag in the absolute magnitude light curves of SE-SNe. \cite{2018A&A...609A.136T} found a correlation between late-time slopes and $\triangle m_{15}$, which is the decay rate from peak upto 15 days post maximum, suggesting that light curves with initial steeper slopes decline faster at later stages.

The $^{56}$Ni mass, ejecta mass and kinetic energy can be measured by modelling the bolometric light curves. \cite{2013MNRAS.434.1098C} estimated median $^{56}$Ni mass between 0.15$-$0.18 M$_{\odot}$ and ejecta masses of 3.9 M$_{\odot}$ and 3.4 M$_{\odot}$ for Type Ib and Type Ic SNe respectively. While \cite{2018A&A...609A.136T} found ejecta masses for SE-SNe to be between 1.1 -- 6.2 M$_{\odot}$, \cite{2019MNRAS.485.1559P} quote the median distribution to be 2.8 $\pm$ 1.5 M$_{\odot}$. Even though high ejecta masses for Type Ic SNe indicates higher progenitor masses, the distribution of ejecta masses for Type Ic SNe hints towards a wide distribution of progenitor masses. Also, the ejecta mass distribution for both Type Ib and Ic SNe are inconsistent with massive single stars and support moderate mass stars in binary associations \citep{2016MNRAS.457..328L}. The most common technique used for direct detection of progenitors involve analysing the  pre-explosion images. So far direct detection has been possible for only one Type Ib SN iPTF13bvn \citep{2013ApJ...775L...7C,2013A&A...558L...1G,2014AJ....148...68B,2014A&A...565A.114F,2015A&A...579A..95K,2016ApJ...825L..22F,2016A&A...593A..68F,2016MNRAS.461L.117E,2017MNRAS.466.3775H,2017MNRAS.469L..94H} and one Type Ic SN 2017ein \citep{2018ApJ...860...90V,2018MNRAS.480.2072K}. The progenitor masses of iPTF13bvn and SN 2017ein were found to be 10$-$20 M$_{\odot}$ \citep{2013ApJ...775L...7C,2013A&A...558L...1G} and 47$-$80 M$_{\odot}$ \citep{2018ApJ...860...90V,2018MNRAS.480.2072K} respectively although the progenitor detection of SN 2017ein is debated \citep{2018ApJ...860...90V,2018MNRAS.480.2072K}.

Metallicity and rotation largely affects the predicted ratios of Type Ib/c SNe, mass-loss rates and the production of WR stars \citep{2008MNRAS.384.1109E}. \cite{2008MNRAS.384.1109E} have shown that stars with high ejecta mass come from massive progenitors. However, if fallback occurs, even less massive stars can eject more mass (for example SN 1999dn, \citealp{2011MNRAS.411.2726B}). \cite{2015A&A...579A..95K} favour low mass binary systems as progenitors of Type Ib/c SNe. On the other hand, \cite{2018ApJ...855..107G} claimed that Type Ic SNe have the most massive progenitors. But, both \cite{2015A&A...579A..95K} and \cite{2018ApJ...855..107G}  suggest the origin of Type Ib/c SNe from young stellar populations. Thus, the exact characterisation of ejecta mass, energy and consecutively the progenitor system remains largely uncertain.

In this paper, we present the photometric and spectroscopic study of a Type Ib SN 2015ap and a Type Ic SN 2016P. The SNe details are given in section \ref{1}. The data acquisition and reduction procedures are described in section \ref{sec:data}. Temporal evolution of the multi-band light curves and colour curves are studied in section \ref{sec:lc} along with the analytical modelling of the bolometric light curve. A detailed description of the spectral evolution and asymmetry in the nebular phase are presented in sections \ref{4} and \ref{5} respectively. Section \ref{6} summarises the main results of this study.

\section{Parameters of SNe}
\label{1}
\subsection{SN 2015ap}
\label{1.1}
SN 2015ap was discovered by Lick Observatory Supernova Search (LOSS) on 2015, September 08 by Ross, Zheng, Filippenko  at R.A.= 02$^{h}$05$^{m}$13.3$^{s}$, Decl.= +06$^{\circ}$06$^{'}$08$^{''}$ (J2000.0). The SN was located 28".7 west and 16".4 south of the center of the host galaxy IC 1776 at a redshift of 0.01138. The classification spectrum was obtained with Copernico 1.82 m telescope (+AFOSC; range 340-820 nm; resolution 1.4 nm) operated by INAF Astronomical observatory of Padova in Asiago, Mount Ekar, Italy, on 2015, September 11 \citep{2015ATel.8039....1T} and the spectrum was found to be similar to those of several broad-lined Type Ib/c SNe close to maximum light. It was re-classified as Type Ib SN using the spectrum obtained on 2015 September 21 by \cite{2015ATel.8081....1S}.
We used the GEneric cLAssification TOol (GELATO) \citep{2008A&A...488..383H} on the first three spectra of SN 2015ap. GELATO compares bins of the spectrum to those of template SNe at similar epochs which suggests that the explosion occurred on 2015 September 05 $\pm$ 1. However, GELATO results will be reasonable only if the light curves also match. We match the $V$-band light curve of SN 2015ap with the light curves of the SNe in our comparison sample (see section \ref{sec:lc}). Performing a template fitting method by applying a simultaneous shift in phase and magnitude, we find that the light curve of SN 2007gr best matches our observed light curve which constrains the explosion epoch as 2015 September 07 $\pm$ 1. To further constrain the explosion epoch, we performed a parabolic fit (described in \cite{2018MNRAS.476.3611G}) to the $V$-band light curve and infer the explosion epoch to be 2015 September 07. Based on the above methods, we conclude that the explosion occurred between 2015 September 04 to 08. We adopt a distance modulus of $\mu$ = 33.27 $\pm$ 0.15 mag for the host galaxy IC 1776 and total A$_{V}$ = 0.115 mag along the line of sight (negligible host extinction) from \cite{2019MNRAS.485.1559P}.
\subsection{SN 2016P}
\label{1.2}
SN 2016P was discovered on 2016 January 19, 21".7 east and 2".8 north of the center of the host galaxy NGC 5374 at R.A. = 13$^{h}$57$^{m}$31.13$^{s}$, Dec = +06$^{\circ}$05$^{'}$51.6$^{''}$ (J2000.0). An optical spectrum obtained with YFOSC at Yunnan Astronomical Observatories on 2015 January 19.9  was found to be consistent with a BL Ic SN a few days before maximum light \citep{2016ATel.8563....1Z}.
Following the same approach as in SN 2015ap, GELATO runs on the first three spectra of SN 2016P estimates the  explosion epoch to be 2016 January 15 $\pm$ 1. The light curve template fitting method gives a best match with SN 2007ru and constrains the explosion epoch to be 2016 January 18 $\pm$ 3. The parabolic fit to the $V$-band light curve upto 10 days estimates the explosion epoch to be around 2016 January 15 $\pm$ 1. The bolometric light curve modelling (see section \ref{sec:abs_bol}) gives the explosion epoch to be 2016 January 15. The results of the above methods constrains the explosion epoch to be between 2016 January 14 $-$ 18 (since discovery was on 2016 January 19, we constrain it upto 2016 January 18). From \cite{2019MNRAS.485.1559P}, we adopt a distance modulus of $\mu$ = 34.19 $\pm$ 0.15 mag for the host galaxy NGC 5374. The galactic {\it E(B-V)} along the line of sight of SN 2016P is 0.024 mag \citep{2011ApJ...737..103S}. Using the equivalent width of the NaID line \cite{2019MNRAS.485.1559P} estimate the host galaxy {\it E(B-V)} = 0.05 $\pm$ 0.02 mag. Due to the poor resolution in our observed spectra, we were unable to measure the equivalent width of the NaID line. We adopt a total extinction of A$_{V}$ = 0.229 $\pm$ 0.092 mag along the line of sight of SN 2016P from \cite{2019MNRAS.485.1559P}.
\section{Data Acquisition and Reduction}
\label{sec:data}
Photometric observing campaigns were carried out from 2015 September 18 to 2016 August 27 for SN 2015ap and from 2016 January 21 to 2016 June 03 for SN 2016P respectively using 1.04m Sampurnanand Telescope (ST); 1.30m Devasthal Fast Optical Telescope (DFOT) located in ARIES, Nainital, India; Copernico 1.82m Telescope operated by INAF Astronomical University of Padova (Mount Ekar), Italy; 2.00m Himalayan Chandra Telescope (HCT), IAO, Hanle, India and 2.54m Nordic Optical Telescope (NOT), La Palma, Canary Island. The imaging observations were done using Johnson-Cousin-Bessel {\it UBVRI} and SDSS {\it ugriz} filters.
Image reduction and photometry were done using IRAF\footnote{Image Reduction and Analysis Facility} and DAOPHOT\footnote{Dominion Astrophysical Observatory Photometry} following the same steps as outlined in \cite{2018MNRAS.476.3611G} and briefly described in Appendix A. The SN magnitudes were calibrated with respect to the local standards (Fig \ref{fig:calibimage_15ap},Tables \ref{tab:optical_observations} and \ref{tab:optical_observations_2016P}). The final magnitudes of SNe 2015ap and 2016P are given in Table \ref{tab:observation_log_2015ap} and Table \ref{tab:observation_log_2016P} respectively.

Low and medium resolution optical spectra of SN 2015ap were obtained with 2.0m HCT and Copernico 1.82m telescope at 14 epochs. For SN 2016P, spectra were obtained at six epochs with HCT. The spectroscopic reduction was done using standard packages in IRAF. The spectra were wavelength and flux calibrated following the steps described in \cite{2018MNRAS.476.3611G}. The log of spectroscopic observations is presented in Table \ref{tab:2015ap_spec_obs} and Table \ref{tab:2016P_spec_obs}.

\section{Temporal evolution of SNe 2015ap and 2016P}
\label{sec:lc}
\begin{table*}
\caption{ Parameters of SNe 2015ap and 2016P  }
\centering
\smallskip
\begin{tabular}{l  c c c c c}
\hline \hline
SN 2015ap                                       & $B$ band     & $V$ band                      & $R$ band              & $I$ band  \\
\hline
 MJD (maximum)                               & 57283.0 $\pm$ 0.5      & 57284.9 $\pm$ 0.5       & 57289.3 $\pm$ 0.5       & 57287.5 $\pm$ 0.5       \\
Magnitude at maximum (mag)                   & 15.54 $\pm$ 0.03       & 15.23 $\pm$ 0.02         & 15.16 $\pm$ 0.02       & 14.76 $\pm$ 0.04    \\
Absolute magnitude at maximum (mag)          & -17.71 $\pm$ 0.19      & -18.04 $\pm$ 0.19       & -18.15 $\pm$ 0.17       & -18.54 $\pm$ 0.16    \\
$\Delta$m$_{15}$ (mag)                       & 1.73 $\pm$ 0.10        & 1.03 $\pm$ 0.07         & 0.74 $\pm$ 0.05         & 0.58 $\pm$ 0.05 \\
{\bf Rise time (rest-frame-day)}             & {\bf 9.9}              & {\bf 11.8}              & {\bf 16.1}              & {\bf 14.3}   \\
\hline \hline
SN 2016P                                        & $B$ band     & $V$ band                      & $R$ band              & $I$ band  \\
\hline
MJD (maximum)                                & 57413.03 $\pm$ 0.5      & 57416.75 $\pm$ 0.5       & 57417.20 $\pm$ 0.5    & 57418.36 $\pm$ 0.5       \\
Magnitude at maximum (mag)                   & 17.37 $\pm$ 0.05       & 16.64 $\pm$ 0.08        & 16.36 $\pm$ 0.08        & 16.15 $\pm$ 0.05    \\
Absolute magnitude at maximum (mag)          & -16.89 $\pm$ 0.12      & -17.53 $\pm$ 0.14       & -17.81 $\pm$ 0.13       & -17.92 $\pm$ 0.11    \\
$\Delta$m$_{15}$ (mag)                       & 1.47 $\pm$ 0.12        & 1.20 $\pm$ 0.06         & 0.93 $\pm$ 0.04         & 0.63 $\pm$ 0.05\\
{\bf Rise time (rest-frame-day)}             & {\bf 6.9}              & {\bf 10.6}              & {\bf 11.0}              & {\bf 12.4}   \\
\hline
\end{tabular}
\newline
\label{tab:params}      
\end{table*}

\begin{figure}
	\begin{center}
		\includegraphics[width=0.5\textwidth]{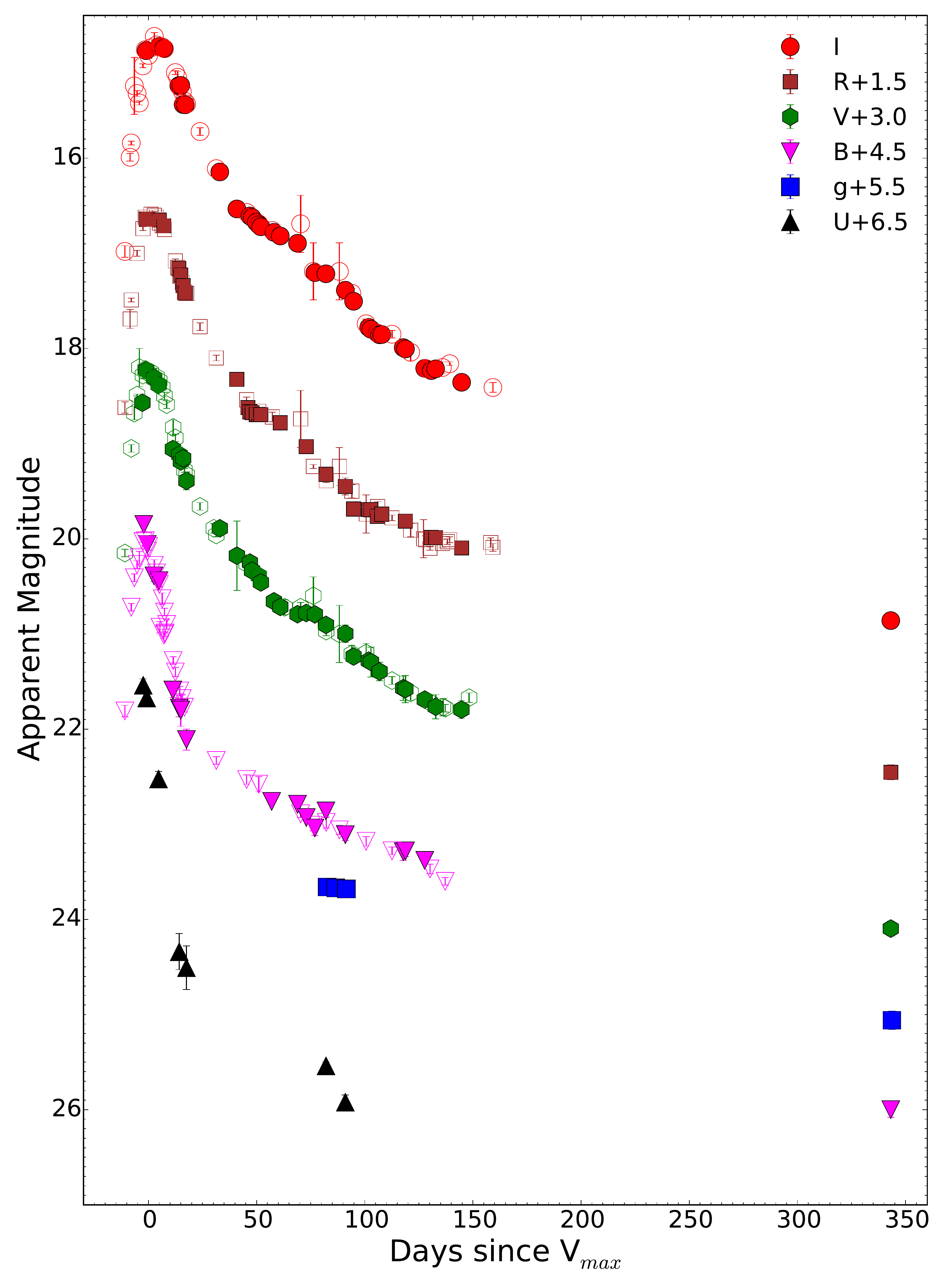}
	\end{center}
	\caption{{\it UgBVRI} light curve evolution of SN 2015ap. The filled symbols represent our data and open symbols represent the data taken from \citealt{2019MNRAS.485.1559P}.}
	\label{fig:lc_2015ap}
\end{figure}

\begin{figure}
	\begin{center}
		\includegraphics[width=0.5\textwidth]{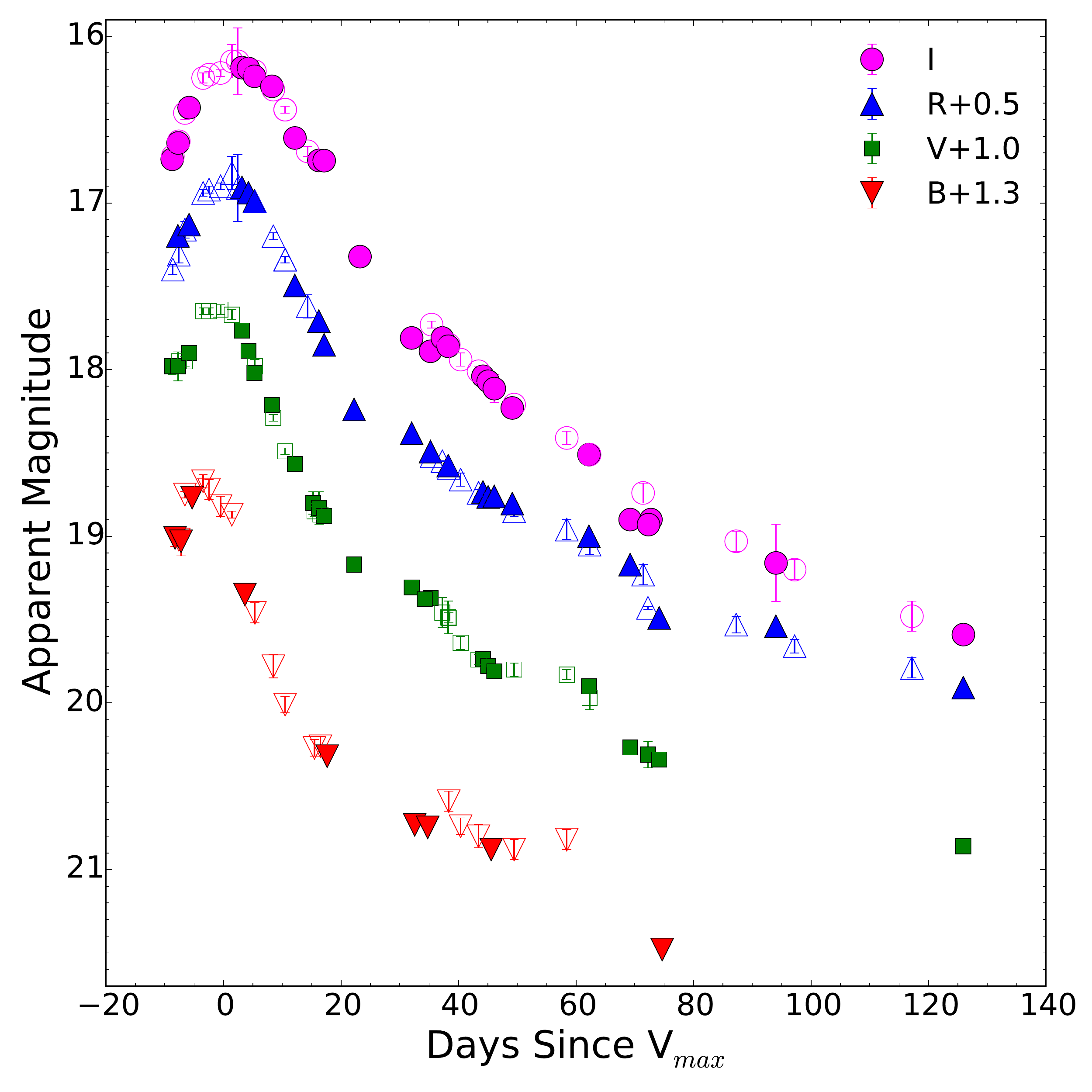}
	\end{center}
	\caption{{\it BVRI} light curve evolution of SN 2016P. The filled symbols represent our data and open symbols represent the data taken from \citealt{2019MNRAS.485.1559P}.}
	\label{fig:lc_2016P}
\end{figure}

The complete multi-band optical light curve of SNe 2015ap and 2016P are shown in Fig \ref{fig:lc_2015ap} and Fig \ref{fig:lc_2016P} respectively. The date of maximum and brightness at peak in different bands were determined by fitting a cubic spline to the {\it BVRI} light curve of both the SNe and are listed in Table \ref{tab:params}. The maximum in $V$-band for SNe 2015ap and 2016P occurred on MJD 57284.86 $\pm$ 0.50 and 57416.75 $\pm$ 0.50 at an apparent magnitude of 15.23 $\pm$ 0.02 mag and 16.64 $\pm$ 0.08 mag respectively. The errors reported are obtained from the interpolated measurements around the peak. We notice that there is a gradual delay in reaching the maximum in the redder passbands. We use days since $V$-maximum as a reference epoch throughout the paper.

\begin{figure*}
	\begin{center}
		\includegraphics[width=1.0\textwidth]{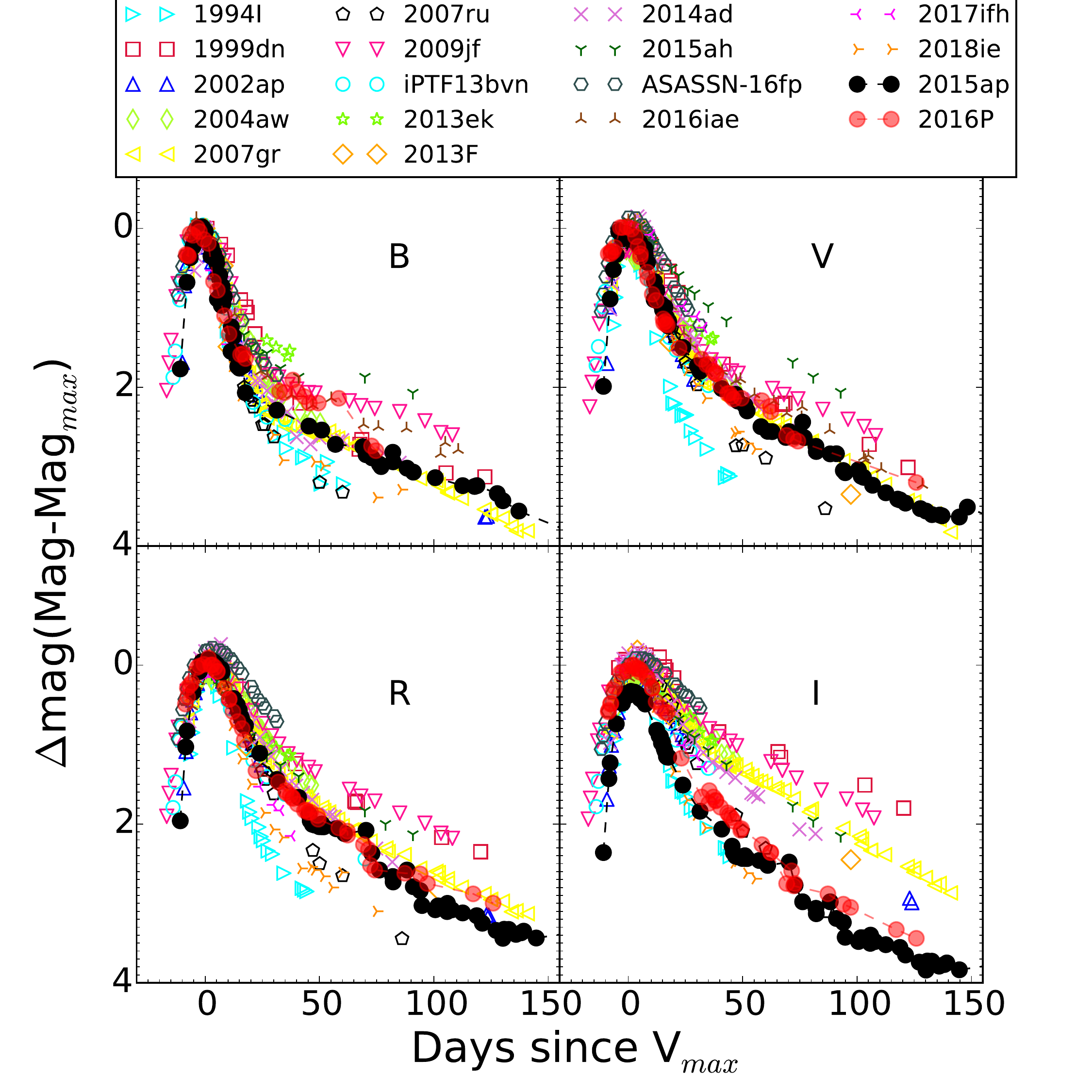}
	\end{center}
	\caption{{\it BVRI/ri} light curve evolution of SN 2015ap and SN 2016P compared with other members of the sub-class. All the light curves are shifted in phase and matched to maximum light.}
	\label{fig:comp_lc}
\end{figure*}

The light curve features of SNe 2015ap and 2016P are compared with those of other well-studied SNe Ib/c in Fig \ref{fig:comp_lc}. We construct a sample of SNe Ib/c from literature and list their parameters in Table \ref{tab:Properties_of_the_comparison_sample}. The comparison sample chosen is vast and diverse. While SNe 2009jf and iPTF13bvn are fast evolving, SNe 1999dn is a prototypical Type Ib SN. Type Ic sample consists of both normal and BL-Ic's to highlight the diversity in this sub-class. We have also included both fast and slow-evolving Type Ic which enables us to make a direct comparison with SN 2016P. In addition to these, we have also selected those Type Ib/c SNe from \cite{2019MNRAS.485.1559P} which have observations in the {\it B} and the {\it V}-bands that are consistent with our observations. This comparison sample represents the properties of the Type Ib/c population.

Fig \ref{fig:comp_lc} shows the light curve evolution up to 150 days post $V$-maximum. The observed SN magnitudes of SNe 2015ap and 2016P along with the comparison sample are normalised with respect to their respective peak magnitudes, and a shift in the time is applied to match the time of maximum. To compare the light curve evolution of Type Ib/c SNe, we estimated decline of the light-curve in the first 15 days after peak ($\triangle m_{15}$) and late phase decline rate between 50$-$100 days. The decline rates are listed in Table \ref{tab:comp_decay_rate}.

\begin{table*}
\caption{Properties of the comparison sample }
\centering
\small\addtolength{\tabcolsep}{-4pt}
\smallskip
\begin{tabular}{l c c c c c c c c c c}
\hline \hline
                       & SNe        & Host galaxy & Distance   & Extinction & SN Type    & Peak Absmag$^\ddagger$ & M$_{Ni}$ & M$_{ej}$ & E$_{k}$  &  Reference$^\dagger$  \\
                       &            &             & (Mpc)      & E(B-V)     &    & $V$-band & M$_{\odot}$     & M$_{\odot}$   & 10$^{52}$ erg  &     \\
\hline
                       & SN 1999dn  & NGC 7714    & 38.5       & 0.100 & Ib  & -16.88   & 0.09 $\pm$ 0.04 & 5.13 $\pm$ 1.82  & 0.37 $\pm$ 0.13 & 2,11,12      \\
		       & SN 2009jf  & NGC 7479    & 33.7       & 0.110 & Ib   & -17.90  & 0.18 $\pm$ 0.04 & 7.34 $\pm$ 1.0 &1.05 $\pm$ 0.15  & 4,11  \\
                       & iPTF 13bvn & NGC 5806    & 23.9       & 0.215 & Ib   & -17.35  & 0.09            & 1.5 $-$ 2.2    & 0.1             & 5     \\
                       & SN 2013ek  & NGC 6984    & 50.1       & 0.037 & Ib   & -17.22  & 0.05 $\pm$ 0.01 & 1.2            & 0.04                & 13 \\
                       & SN 2015ah  & UGC12295    & 58.9       & 0.091 & Ib   & -17.53  & 0.092 $\pm$ 0.07 & 2             & 0.09                & 13 
\\
\hline
		       & SN 1994I   & M51         & 6.3        & 0.452 & Ic  & -17.49   & 0.06 $\pm$ 0.01 & 0.72 $\pm$ 0.04 & 0.07 $\pm$ 0.01  & 1,11,12     \\
                       & SN 2002ap  & NGC 0628    & 9.9        & 0.090 & Ic-BL & -18.11 & 0.09            & $\geq$ 1.5 & ---                & 6      \\
                       & SN 2004aw  & NGC 3997    & 72.8       & 0.092 & Ic &  -17.30 & 0.22 $\pm$ 0.08 & 6.49 $\pm$ 2.32  & 0.90 $\pm$ 0.32 & 7,11,12      \\
                       & SN 2007gr  & NGC 1058    & 9.7        & 0.092 & Ic & -17.11  & 0.04 $\pm$ 0.01 & 1.70 $\pm$ 0.23 & 0.08 $\pm$ 0.01 & 3,11,12     \\
                       & SN 2007ru  & UGC 12381   & 64.2       & 0.270 & Ic-BL & -18.92 & 0.34 $\pm$ 0.09 & 3.59 $\pm$ 1.52 & 1.43 $\pm$ 0.57  & 8,11,12     \\
                       & SN 2013F   & IC 5325     & 17.6       & 1.418 & Ic   & -18.55  & 0.15            & 1.4            & 0.11           & 13 \\         
                       & SN 2014ad  & MRK 1309    & 26.6       & 0.220 & Ic-BL & -18.72 &  0.24           & 3.3 $\pm$ 0.8 & 1 $\pm$ 0.3                  & 9,12     \\
                       & ASASSN-16fp& NGC 5806    & 23.9       & 0.074 & Ic &  -18.36 & 0.1                & 4.5 $\pm$ 0.3 & 0.69                & 10,12     \\
                       & SN 2016iae & NGC 1532    & 17.3       & 0.664 & Ic &  -18.23 & 0.13                & 2.2  & 0.18               & 13     \\
                       & SN 2017ifh & 2MASX J06350438+5026278 &  156.3 & 0.167 & Ic-BL & -18.55 & 0.16 $\pm$ 0.01  & 3.0  &   0.77           & 13 \\
                       & SN 2018ie  & NGC 3456    & 46.8         & 0.08    & Ic-BL      & -16.94  & 0.026 $\pm$ 0.002    & 1.5  & 0.79             & 13 \\ 
\hline		                                                                                         
\end{tabular}
\newline
$^\dagger$ REFERENCES.-- (1)\cite{1996AJ....111..327R}; (2)\cite{2011MNRAS.411.2726B}; (3)\cite{2009A&A...508..371H}; (4)\cite{2011MNRAS.413.2583S}, NED; (5)\cite{2014MNRAS.445.1932S}, NED; (6)\cite{2003ApJ...592..467Y,2003PASP..115.1220F}, NED; (7)\cite{2006MNRAS.371.1459T}, NED; (8)\cite{2009ApJ...697..676S}; (9)\cite{2018MNRAS.475.2591S}; (10)\cite{2018MNRAS.473.3776K}; (11)\cite{2013MNRAS.434.1098C}; (12) NED; (13)\cite{2019MNRAS.485.1559P}. 
$^\ddagger$ Values are scaled with respect to H$_{o}$ = 73 km sec$^{-1}$ Mpc$^{-1}$.
\label{tab:Properties_of_the_comparison_sample}        
\end{table*}

\cite{2011ApJ...741...97D} states that the average value of $\triangle m_{15}(V)$ and $\triangle m_{15}(R)$ from a literature sample of Type Ib/c SNe are 0.94 $\pm$ 0.31 mag and 0.74 $\pm$ 0.21 mag respectively. For an individual Type Ib, Ic and Ic-BL sample, \cite{2011ApJ...741...97D} estimate $\triangle m_{15}(R)$ to be 0.7 $\pm$ 0.1 mag, 0.9 $\pm$ 0.5 mag and 0.7 $\pm$ 0.1 mag. The $\triangle m_{15}(V)$ values of SN 2015ap matches well with iPTF13bvn, is higher than the value quoted in \cite{2011ApJ...741...97D} but is consistent with the values reported in \cite{2018A&A...609A.136T}. The $\triangle m_{15}$ values in SN 2016P are in between those of the Type Ic comparison sample used in this work, however they are a bit faster than the average values reported in \cite{2011ApJ...741...97D} and \cite{2018A&A...609A.136T}. The late-time light curves are faster than the average decay rate of $^{56}$Co to $^{56}$Fe. \cite{2018A&A...609A.136T} estimated the late time decay rates between 1.4 -- 1.8 mag (100 days)$^{-1}$ and 1.7 -- 2.7 mag (100 days)$^{-1}$ for Type Ib and Type Ic sample respectively. Table \ref{tab:comp_decay_rate} and Fig \ref{fig:comp_lc}, thus, shows that both SNe 2015ap and 2016P shows faster evolution than the $^{56}$Co -- $^{56}$Fe decay rates. This implies inefficient $\gamma$-ray trapping for both the SNe.

\begin{table*}
\begin{minipage}{400mm}
\caption{Early and late time decay rates of a sample of Type Ib/c SNe.}
\label{tab:comp_decay_rate}
\begin{tabular}{ccc|ccccc}
\hline \hline
\multicolumn{1}{c}{Supernova}
&\multicolumn{2}{c}{$\triangle m_{15}$ (mag)}
&\multicolumn{4}{c}{Late time (50-100 days) decay rates (mag (100 days)$^{-1}$)}
&\multicolumn{1}{c}{Reference} \\
\hline
                      & {$V$}&{$R$} & {$B$}&{$V$}&{$R$}&{$I$}& \\
\hline
SN 1999dn             & 0.49 $\pm$ 0.08      & 0.31 $\pm$ 0.03       & 1.07 $\pm$ 0.19   & 1.54 $\pm$ 0.09      & 1.32 $\pm$ 0.09     & 1.15 $\pm$ 0.07     & 3 \\       
SN 2007gr             & 1.31 $\pm$ 0.05      & 0.46 $\pm$ 0.07       & 1.11 $\pm$ 0.06   & 1.70 $\pm$ 0.05      & 1.94 $\pm$ 0.07     & 1.64 $\pm$ 0.04   & 3,1  \\
SN 2009jf             & 0.50                 & 0.31                  & 0.89 $\pm$ 0.14   & 1.36 $\pm$ 0.07      & 1.61 $\pm$ 0.02     & 1.71 $\pm$ 0.11     & 4,5\\
iPTF13bvn             & 1.01                 & 0.90                  & ---               & ---                  & ---                 & ---       & 10   \\ 
SN 2013ek                & 0.78 $\pm$ 0.08      & 0.60 $\pm$ 0.10       & ---               & ---                  & ---                 & ---       & 12,3 \\   
SN 2015ah                & 0.83 $\pm$ 0.04      & 0.71 $\pm$ 0.01       & ---               & ---                  & ---                 & ---       & 12,3 \\
{\bf SN 2015ap}	      & 1.03 $\pm$ 0.07      & 0.74 $\pm$ 0.05       & 1.10 $\pm$ 0.15   & 1.55 $\pm$ 0.14      & 2.14 $\pm$ 0.18     & 1.88 $\pm$ 0.17   & 3  \\ \\   

SN 1994I              & 1.74 $\pm$ 0.02      & 1.46 $\pm$ 0.02       & ---               & 1.84 $\pm$ 0.05      & ---               & ---                 & 2 \\                                          
SN 2002ap             & 1.85 $\pm$ 0.45      & ---                   & 1.70 $\pm$ 0.04   & 2.10 $\pm$ 0.02      & 1.60 $\pm$ 0.01     & 1.70 $\pm$ 0.01    & 3,6 \\       
SN 2004aw             & 0.62 $\pm$ 0.03      & 0.41 $\pm$ 0.03       & 1.35 $\pm$ 0.18   & 1.74 $\pm$ 0.22      & 1.36 $\pm$ 0.14     & 1.53 $\pm$ 0.19   & 5  \\
SN 2007ru             & 0.92                 & 0.69                  & ---               & 2.1                  & 2.8                 & 3.0           & 7   \\
SN 2013F              & 1.43 $\pm$ 0.26      &  1.05 $\pm$ 0.28      & ---               & ---                  & ---                 & ---       & 12,3 \\
SN 2014ad             & 0.95 $\pm$ 0.06      & 0.77 $\pm$ 0.03       & 0.89 $\pm$ 0.10   & 2.19 $\pm$ 0.13      & 2.17 $\pm$ 0.05     & 2.04 $\pm$ 0.11     & 3,8 \\
ASASSN-16fp           & 0.60 $\pm$ 0.02      & 0.42 $\pm$ 0.02       & ---               & ---                  & ---                 & ---     & 3,9 \\
SN 2016iae            & 0.91 $\pm$ 0.25      & ---                   & 1.29 $\pm$ 0.24   & 1.61 $\pm$ 0.19      & ---                 & ---     & 12,3 \\
SN 2018ie             & 1.41 $\pm$ 0.16      & 1.06 $\pm$ 0.13       & ---               & ---                  & ---                 & ---     & 12,3\\
{\bf SN 2016P}        & 1.20 $\pm$ 0.06      & 0.93 $\pm$ 0.04       & ---               & 1.75 $\pm$ 0.22      & 1.78 $\pm$ 0.18     & 2.09 $\pm$ 0.21     & 3 \\\\

{\bf Type Ib/c, Ic-BL} & 0.94 $\pm$ 0.31     & 0.74 $\pm$ 0.21       & ---               &  ---                 & ---                 & ---    & 11 \\
{\bf Type Ib SNe}      & 1.03 $\pm$ 0.19     & 0.75 $\pm$ 0.21       & ---               &  ---                 & ---                 & ---    & 13 \\
{\bf Type Ic SNe}      & 0.90 $\pm$ 0.22     & 0.62 $\pm$ 0.24        & ---               &  ---                 & ---                 & ---    & 13 \\  
\hline 
\end{tabular}
\newline
$^\dagger$ REFERENCES.--(1) \cite{2009A&A...508..371H}; (2) \cite{1996AJ....111..327R}; (3) This work ; (4) \cite{2006MNRAS.371.1459T}; 
\newline
(5) \cite{2011MNRAS.416.3138V,2011MNRAS.413.2583S}; (6) \cite{2009A&A...508..371H}; (7) \cite{2009ApJ...697..676S}; (8) \cite{2018MNRAS.475.2591S}; 
\newline
(9) \cite{2018MNRAS.473.3776K}; (10) \cite{2014MNRAS.445.1932S};  (11) \cite{2011ApJ...741...97D}; (12) \cite{2019MNRAS.485.1559P}; (13) \cite{2018A&A...609A.136T}
\end{minipage}
\end{table*}

\subsection{Colour evolution of SNe 2015ap and 2016P}
\label{sec:colour}
\begin{figure}
	\begin{center}
		\includegraphics[width=1.0\columnwidth]{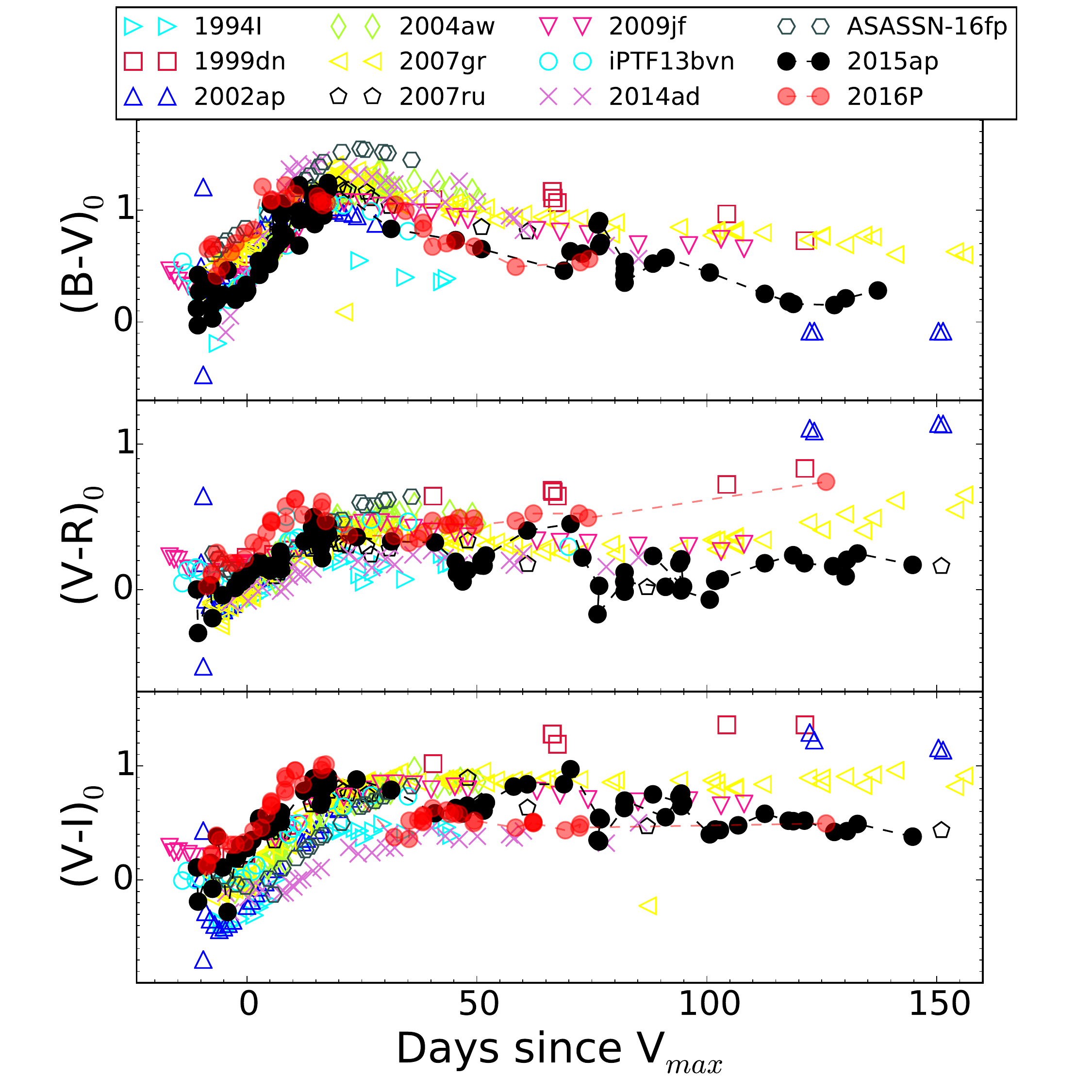}
	\end{center}
	\caption{A comparison of the {\it (B-V)$_{0}$, (V-R)$_{0}$} and {\it (V-I)$_{0}$} colours of SNe 2015ap and 2016P with other representative SNe Ib/c. The colours are corrected for the reddening values given in Table \ref{tab:Properties_of_the_comparison_sample}.}
	\label{fig:colourcurve}
\end{figure}
The colour evolution of SNe 2015ap and 2016P along with other members of the comparison sample are displayed in Fig \ref{fig:colourcurve}. A diversity exists among the colour variation due to expansion of photospheric envelope and the changing temperatures. All the colour curves have been dereddened by the value reported in Table \ref{tab:Properties_of_the_comparison_sample}. The pre-maximum {\it $(B-V)_{o}$} colours of all the SNe are similar and bluer which occurs due to the increase in photospheric temperatures as SN brightens. From $-$11 day to 2 day, the {\it $(B-V)_{o}$} colours of SN 2015ap increases from 0.12 mag to 0.61 mag. During the maximum light, the {\it $(B-V)_{o}$} colour remains around 1.1 mag and becomes bluer post 100 day. The {\it $(B-V)_{o}$} colour evolution of SN 2015ap is similar to all the SNe of Type Ib comparison sample. A similar trend is noticed for the {\it $(V-R)_{o}$} colours of SN 2015ap which is blue ($\sim$ 0.03 mag) at pre-maximum and becomes redder with time. \cite{2011ApJ...741...97D} found that the average {\it $(V-R)_{o}$} colour 10 days post $V$-maximum should be around 0.26 $\pm$ 0.10 mag for Type Ib/c SNe, if the estimated extinction are well within errors. The {\it $(V-R)_{o}$} colour of SN 2015ap is between 0.25 mag to 0.35 mag around 10 days post $V$-maximum which is well in agreement with the values of \cite{2011ApJ...741...97D}. A similar trend is noticed for the {\it $(V-I)_{o}$} colours of SN 2015ap which becomes redder as we move to maximum and remains, constant thereafter.

The {\it $(B-V)_{o}$} colour evolution of SN 2016P increases from 0.6 mag to 1.2 mag 10 days after $V$-maximum. The {\it $(V-R)_{o}$} colours of SN 2016P becomes redder increasing from 0.2 to 0.6 mag. \cite{2016MNRAS.458.1618D} and \cite{2011ApJ...741...97D} have found that the {\it $(V-R)_{o}$} colours 10 days post $V$-maximum shows an average scatter ranging between 0.18 to 0.34 mag. At similar epochs, our value of {\it $(V-R)_{o}$} $\sim$ 0.62 mag indicates the SN to be intrinsically redder. The {\it $(V-I)_{o}$} colours of SN 2016P follows similar trend as {\it $(V-R)_{o}$} which increases upto 1 mag. Post 100 days, the {\it $(V-I)_{o}$} colour becomes blue. SN 2016P is redder than most of the Type Ic SNe comparison sample. This indicates either presence of absorbers in the system or the SN is intrinsically red.

\subsection{Absolute magnitude and Bolometric light curves}
\label{sec:abs_bol}
\begin{figure}
	\begin{center}
		\includegraphics[width=0.5\textwidth]{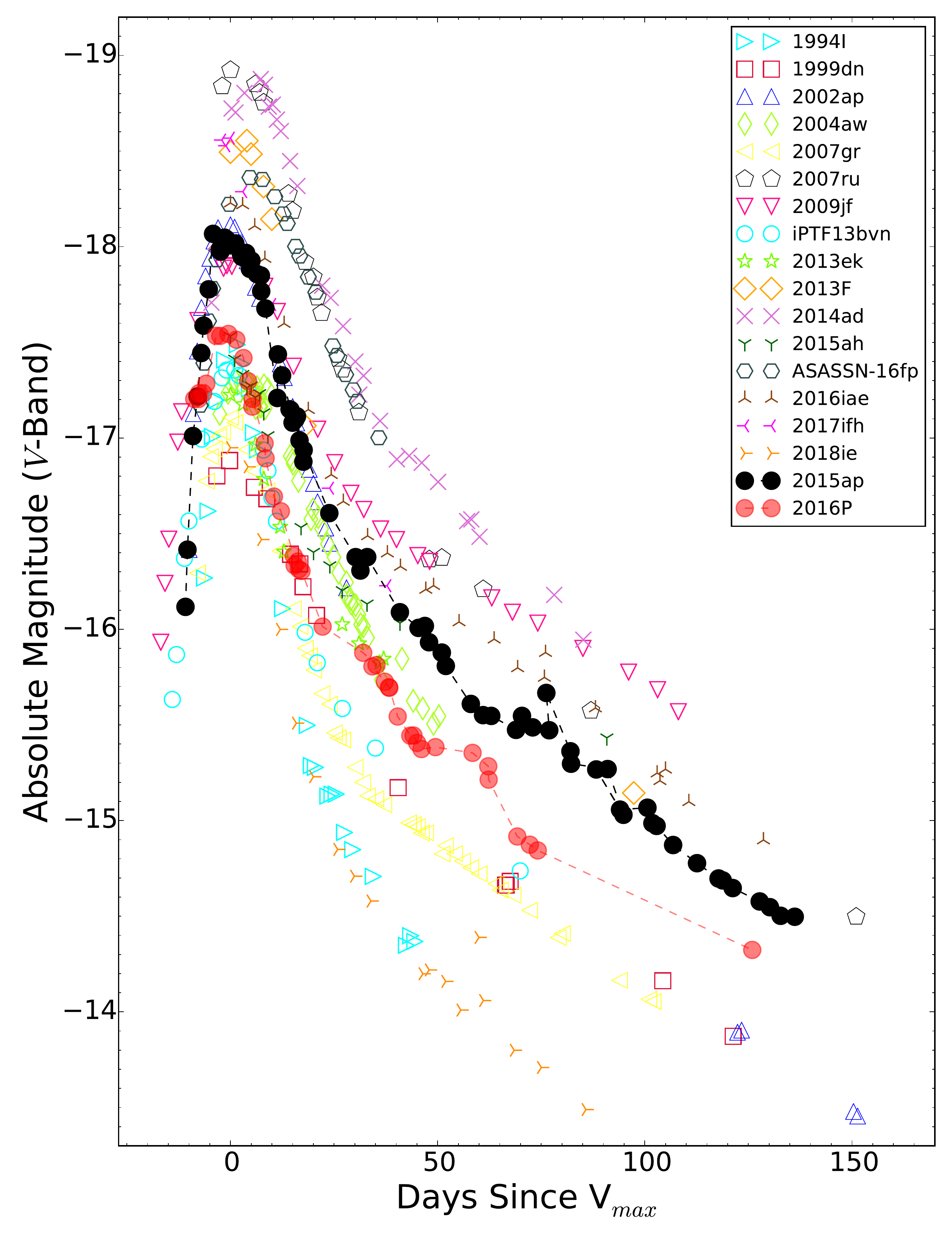}
	\end{center}
	\caption{Absolute magnitude $V$-band light curves of SNe 2015ap and 2016P compared to other members of the sub-class. The light-curves have been corrected for distance and extinction given in Table \ref{tab:Properties_of_the_comparison_sample}. }
	\label{fig:absmag}
\end{figure}
Figure \ref{fig:absmag} shows the absolute magnitude light curves of SNe 2015ap and 2016P along with other members of the sub-class which are corrected for extinction and reddening values reported in Table \ref{tab:Properties_of_the_comparison_sample}. The peak absolute magnitudes of SNe 2015ap and 2016P are listed in Table \ref{tab:params}. The absolute magnitudes of SN 2015ap in $V$ and $R$-bands are $-$18.04 $\pm$ 0.19 mag and $-$18.15 $\pm$ 0.17 mag, respectively, which are comparable with the average distribution of Type Ib SNe ($-$17.98 $\pm$ 0.13 mag; \citealp{2006AJ....131.2233R}, $-$17.9 $\pm$ 0.9 mag; \citealp{2011ApJ...741...97D}). As compared with the recent results by  \cite{2018A&A...609A.136T}, the average value of Type Ib SNe to be $-$17.07 $\pm$ 0.56, $V$-band absolute magnitude of SN 2015ap is brighter by $\sim$ 0.5 mag. The $V$-band absolute magnitude light curve of SN 2015ap shows that it is one of the bright SN among the Type Ib SNe used for comparison. The $R$-band absolute magnitude of SN 2016P is $\sim$ 0.5 mag fainter than the sample of Type Ic SNe ($-$18.3 $\pm$ 0.6; \citealp{2011ApJ...741...97D}). The $V$-band absolute magnitude is also fainter than the sample of Type Ic by 1 mag ($-$18.51 $\pm$ 0.10 mag; \citealp{2006AJ....131.2233R}) while the values are consistent with the recent results by \cite{2018A&A...609A.136T}. Table \ref{tab:Properties_of_the_comparison_sample} shows that SN 2015ap lies among the brighter end of the Type Ib comparison sample while SN 2016P is of average brightness object among the Type Ic comparison sample.

We constructed the bolometric light curves of SNe 2015ap and 2016P using black-body (BB) fitting model. The bolometric lightcurves were constructed by using the photometric data of at least 3 filters to fit black-body spectral energy distribution (SED). For SN 2015ap, however, we also used the pre-peak data of 2 filters since we need pre-peak bolometric lightcurve to pose more stringent constraint on the explosion epoch. Neglecting the dilution effect of the ejecta \citep{2012MNRAS.424.2139D} and assuming that the SN emission is mostly blackbody, we estimated the blackbody temperature (T) and photospheric radius (R$_{\rm ph}$) as free parameters of our code. We neglect the blue-ultraviolet (UV) suppression that yields lower UV and bright optical luminosities. Using the values of T and R$_{\rm ph}$, obtained through BB fitting and considering the contribution of far-UV and far-IR emission, we estimated the bolometric luminosities using the formula 4$\pi$R$_{\rm ph}^2$$\sigma$T$^{4}$. However, in reality, the spectra of SNe Ib/c are not black bodies and there is substantial deviation from this at early and late times due to line blanketing and line emission/loss of photosphere respectively. With the assumption that the photospheric velocity (v$_{\rm ph}$) is equal to the scale velocity (v$_{\rm sc}$), we have modelled the light curves of both the SNe. The photospheric velocity values are taken from the SYN++ modelling of the spectrum done in Sec \ref{4}. We have constrained the priors in the code to obtain physical values of the obtained parameters.
In order to fit the bolometric light curves of SNe 2015ap and 2016P, we used Markov Chain Monte-Carlo (MCMC) simulations on the light curves to obtain the best-fit parameters. The optical opacity is fixed to be 0.07 cm$^{2}$ g$^{-1}$ which is justified if electron scattering opacity is the dominant source (see for example \citealt{2018A&A...609A.136T}).

\begin{figure*}
	\begin{center}
		\includegraphics[width=0.45\textwidth,angle=0]{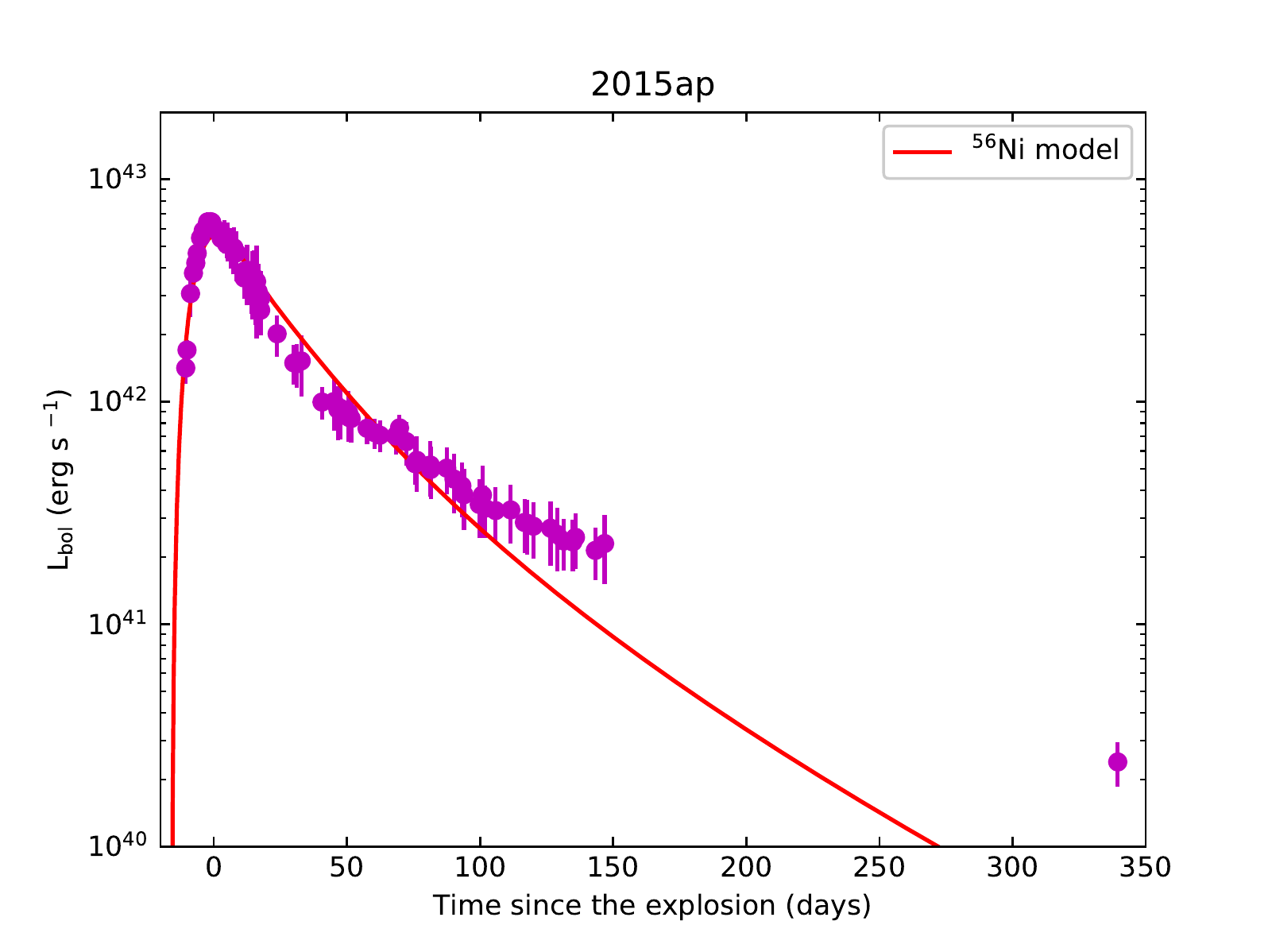}
        \includegraphics[width=0.45\textwidth,angle=0]{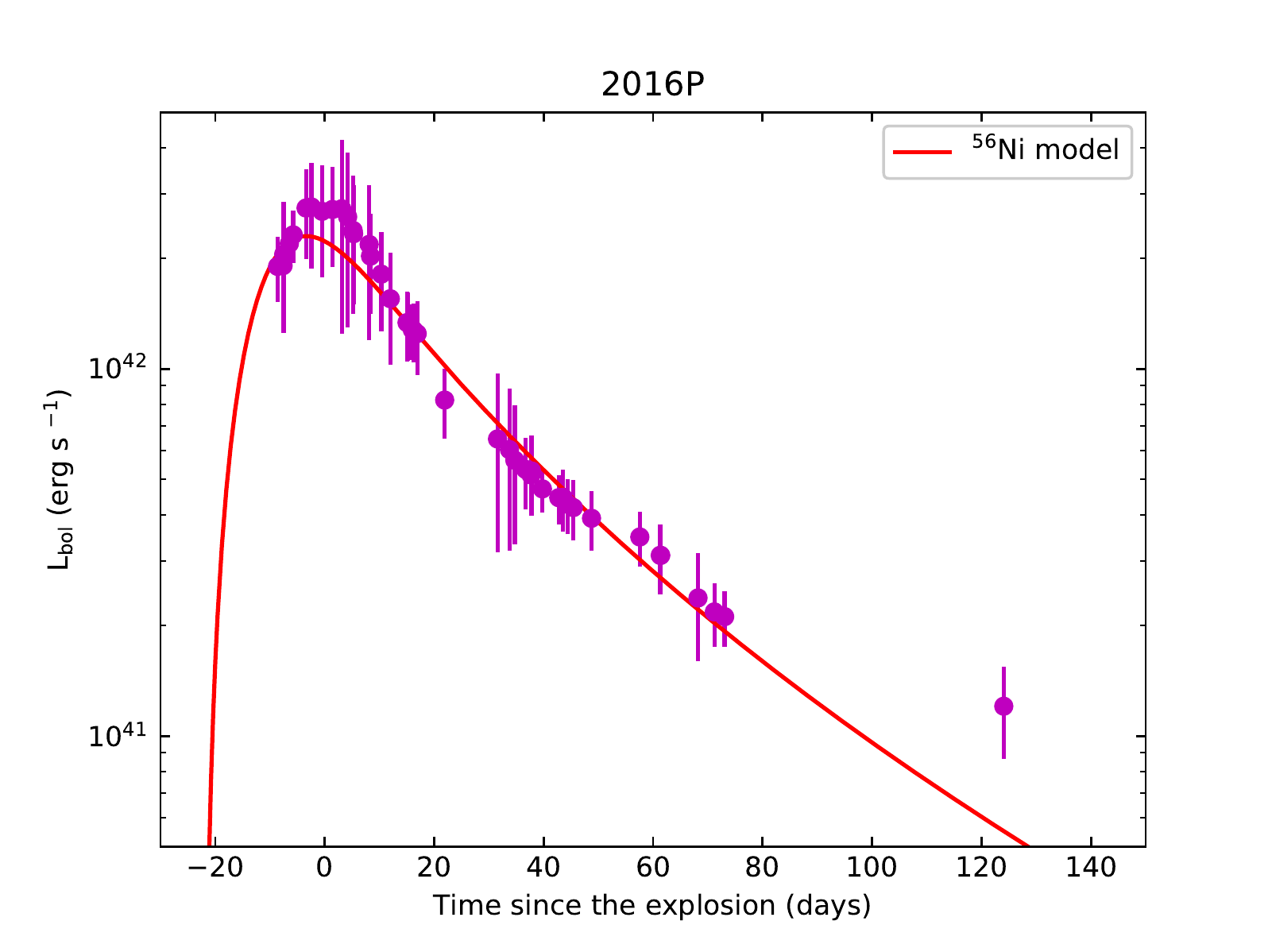}
	\end{center}
	\caption{The bolometric light curves of SNe 2015ap (the \textbf{left} panel) and 2016P (the \textbf{right} panel) reproduced by the $^{56}$Ni model. The abscissa represents time since the explosion in rest frame. The open circles are derived using two band data while filled circles are derived using three or four band data.}
	\label{ni-fits}
\end{figure*}

\begin{figure*}
	\begin{center}
		\includegraphics[width=0.8\textwidth,angle=0]{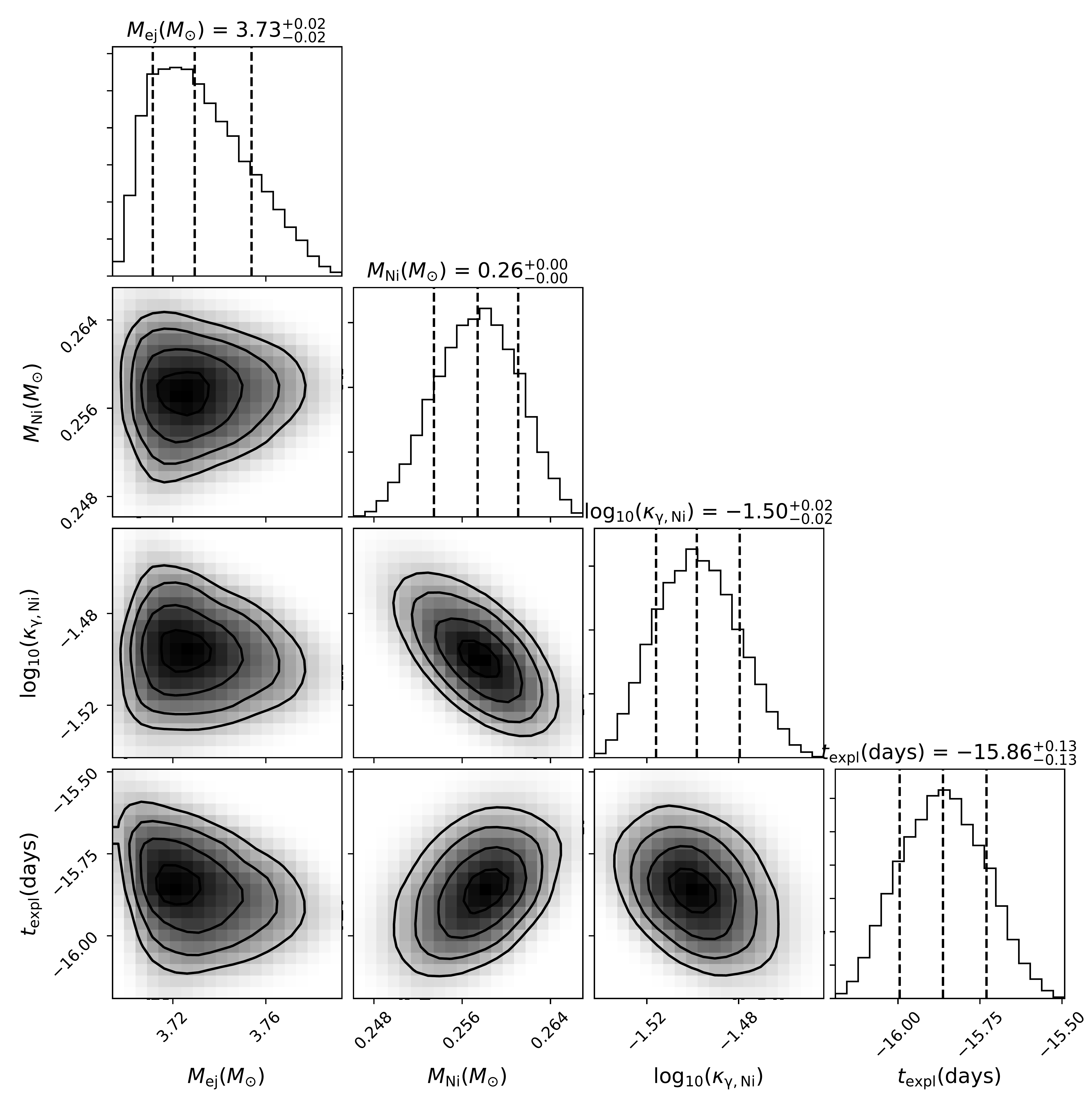}
	\end{center}
	\caption{Corner plot of SN 2015ap showing the parameters (M$_{\rm ej}$, M$_{\rm Ni}$, $\kappa_{\rm \gamma}$, t$_{\rm expl}$) of the $^{56}$Ni model.}
	\label{corner1}
\end{figure*}

\begin{figure*}
	\begin{center}
		\includegraphics[width=0.8\textwidth,angle=0]{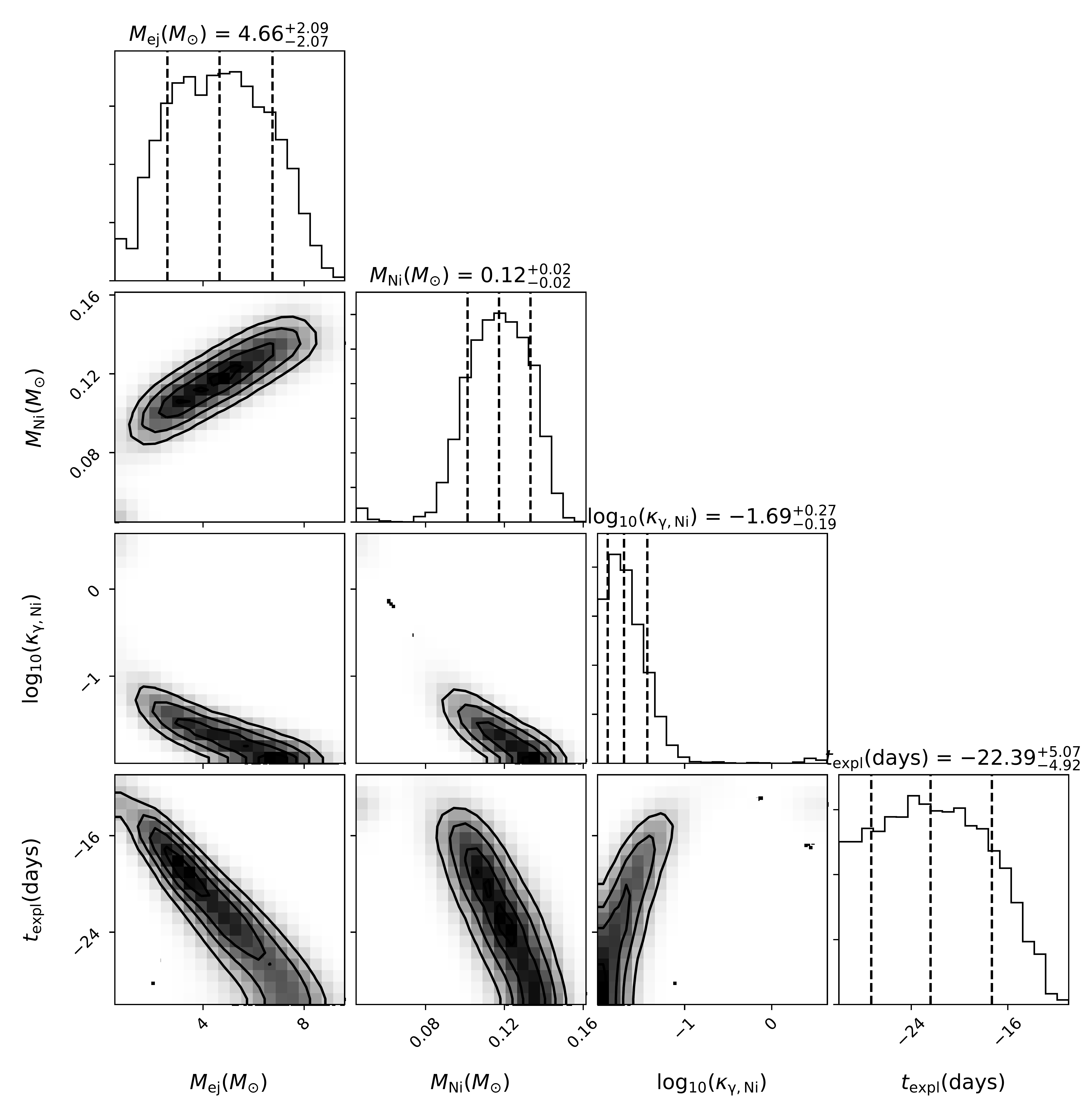}
	\end{center}
	\caption{Corner plot of SN 2016P showing the parameters (M$_{\rm ej}$, M$_{\rm Ni}$, $\kappa_{\rm \gamma}$, t$_{\rm expl}$) of the $^{56}$Ni model.}
	\label{corner}
\end{figure*}

{\bf $^{56}$Ni model:} With the assumption that the bolometric luminosity is powered by the decay of $^{56}$Ni to $^{56}$Co and $^{56}$Co to $^{56}$Fe, we fit the bolometric light curves using the $^{56}$Ni model. The $^{56}$Ni model was constructed taking into account the assumptions of \cite{1982ApJ...253..785A} and \cite{2012ApJ...746..121C}. The employed free parameters of $^{56}$Ni model are the ejecta mass M$_{\rm ej}$, $^{56}$Ni mass M$_{\rm Ni}$, gamma-ray opacity of $^{56}$Ni decay photons $\kappa$$_{\rm \gamma,Ni}$ and explosion time t$_{\rm expl}$.
Following \cite{1982ApJ...253..785A}, $\tau_{m}$ determines the width of the bolometric light curve and can be expressed in terms of opacity ($\kappa$), ejecta mass M$_{\rm ej}$, and the photospheric velocity at luminosity peak v$_{\rm ph}$:
	\begin{equation}
	\tau_{m} = \sqrt{2}{\bigg(\frac{k}{\beta
			c}\bigg)^{\frac{1}{2}}}{\bigg(\frac{M_{\rm ej}}{v_{\rm ph}}\bigg)^{\frac{1}{2}}}
	\end{equation}
	\noindent
	where $\beta=13.8$ is a constant of integration \citep{1982ApJ...253..785A} and {\it c} is the speed of light.
	The kinetic energy for spherically symmetric ejecta with uniform density is:
	\begin{equation}
	E_{k} = \frac{3}{10} M_{\rm ej} v_{\rm ph}^{2}
	\end{equation}
	\noindent
The best-fit parameters of the model are tabulated in Table \ref{tab:para} and the best-fit $^{56}$Ni model of SNe 2015ap and 2016P are displayed in Fig \ref{ni-fits}. The associated corner plots showing confidence level of different parameters are shown in Fig \ref{corner1} and Fig \ref{corner}, respectively. For SN 2015ap, the $^{56}$Ni model is able to reproduce the peak luminosity however it underestimates the luminosity at late times.  Similarly, for SN 2016P the $^{56}$Ni model fails to reproduce the luminosity around peak and is a good match to the observed data at late times. In both the SNe, this indicates that $^{56}$Ni may not alone be the powering source and an additional source of energy is required to generate the observed luminosity.

\begin{table*}
\caption{Parameters of the various models. The uncertainties are 1$\sigma$.}
\label{tab:para}
\begin{center}
{\scriptsize
\begin{tabular}{ccccccccccccc}
\hline\hline
	&	$\kappa$	&	$v_{\mathrm{sc}}$	&	$M_{\mathrm{ej}}$	&	$M_{\mathrm{Ni}}$	&	$P_0$	&	$B_p$	&	$\kappa_{\gamma,\mathrm{Ni}}$	&	$\kappa_{\gamma,\mathrm{mag}}$	&	$t_\mathrm{expl}$$^{\star}$	&	$\chi^2/\mathrm{dof}$	\\
	&	(cm$^2$ g$^{-1}$)	&	($10^9$cm s$^{-1}$)	&	(M$_{\odot}$)	&	(M$_{\odot}$)	&	(ms)	&	($10^{14}$~G)	&	(cm$^2$ g$^{-1}$)	&	(cm$^2$ g$^{-1}$)	&	(days)	&		\\
\hline
\hline
{\bf SN 2015ap}\\
\hline
$^{56}$Ni	&	0.07	&	1.6	&	$3.73^{+0.02}_{-0.02}$	&	$0.26^{+0.00}_{-0.00}$	&	-	&	-	&	$0.032^{+0.001}_{-0.001}$	&	-	&	$-15.86^{+0.13}_{-0.13}$	&	113.64/72	\\
magnetar+$^{56}$Ni	&	0.07	&	1.6	&	$3.90^{+0.21}_{-0.15}$	&	$0.01^{+0.00}_{-0.00}$	&	$25.85^{+0.50}_{-0.67}$	&	$28.39^{+1.64}_{-1.66}$	&	$0.26^{+0.05}_{-0.04}$	&	$14.26^{+24.59}_{-8.49}$	&	$-13.17^{+0.20}_{-0.21}$	&	10.87/69	\\
\hline
{\bf SN 2016P}\\
\hline
$^{56}$Ni	&	0.07	&	1.5	&	$4.66^{+2.09}_{-2.07}$	&	$0.12^{+0.02}_{-0.02}$	&	-	&	-	&	$0.02^{+0.02}_{-0.01}$	&	-	&	$-22.39^{+5.07}_{-4.92}$	&	12.57/38	\\
magnetar+$^{56}$Ni	&	0.07	&	1.5	&	$5.13^{+1.03}_{-1.03}$	&	$0.002^{+0.006}_{-0.004}$	&	$36.51^{+2.16}_{-2.31}$	&	$35.30^{+5.25}_{-3.60}$	&	$1.82^{+17.94}_{-1.66}$	&	$1.19^{+13.09}_{-1.11}$	&	$-16.31^{+1.32}_{-1.47}$	&	3.18/35	\\
\hline\hline
\end{tabular}}
\end{center}
\par
{$\star$ The value of $t_\mathrm{expl}$ is with respect to the date of the first $R$-band observation; the lower limit is set to be $-30$ here. \newline}
\end{table*}

{\bf $^{56}$Ni + Magnetar model:} One of the most prevailing model adopted to model the light curves of SNe that cannot be explained by the $^{56}$Ni model is the magnetar model \citep{2013ApJ...770..128I,2015ApJ...799..107W}. However, as pointed out by \cite{2013ARA&A..51..457N}, the core-collapse SNe can yield $\leq$ 0.2 M$_{\odot}$ of $^{56}$Ni and the luminosity produced by $^{56}$Ni decay cannot be neglected in modeling the SNe whose peak luminosities are not very high. Therefore, a hybrid model taking into account the contribution from a newly born magnetar and a moderate amount of $^{56}$Ni could be a promising model for SNe 2015ap and 2016P. We try to fit the light curves of both SNe 2015ap and 2016P using the $^{56}$Ni + Magnetar model \citep{2015ApJ...807..147W}. The free parameters of the $^{56}$Ni + magnetar model are $\kappa$, M$_{\rm ej}$, magnetic field B$_{p}$, the initial rotational period P$_{0}$, the gamma-ray opacity of magnetar photons $\kappa$$_{\gamma, mag}$,  the gamma-ray opacity of $^{56}$Ni cascade decay photons $\kappa$$_{\gamma, mag}$, and t$_{\rm expl}$.

\begin{figure*}
	\begin{center}
		\includegraphics[width=0.45\textwidth,angle=0]{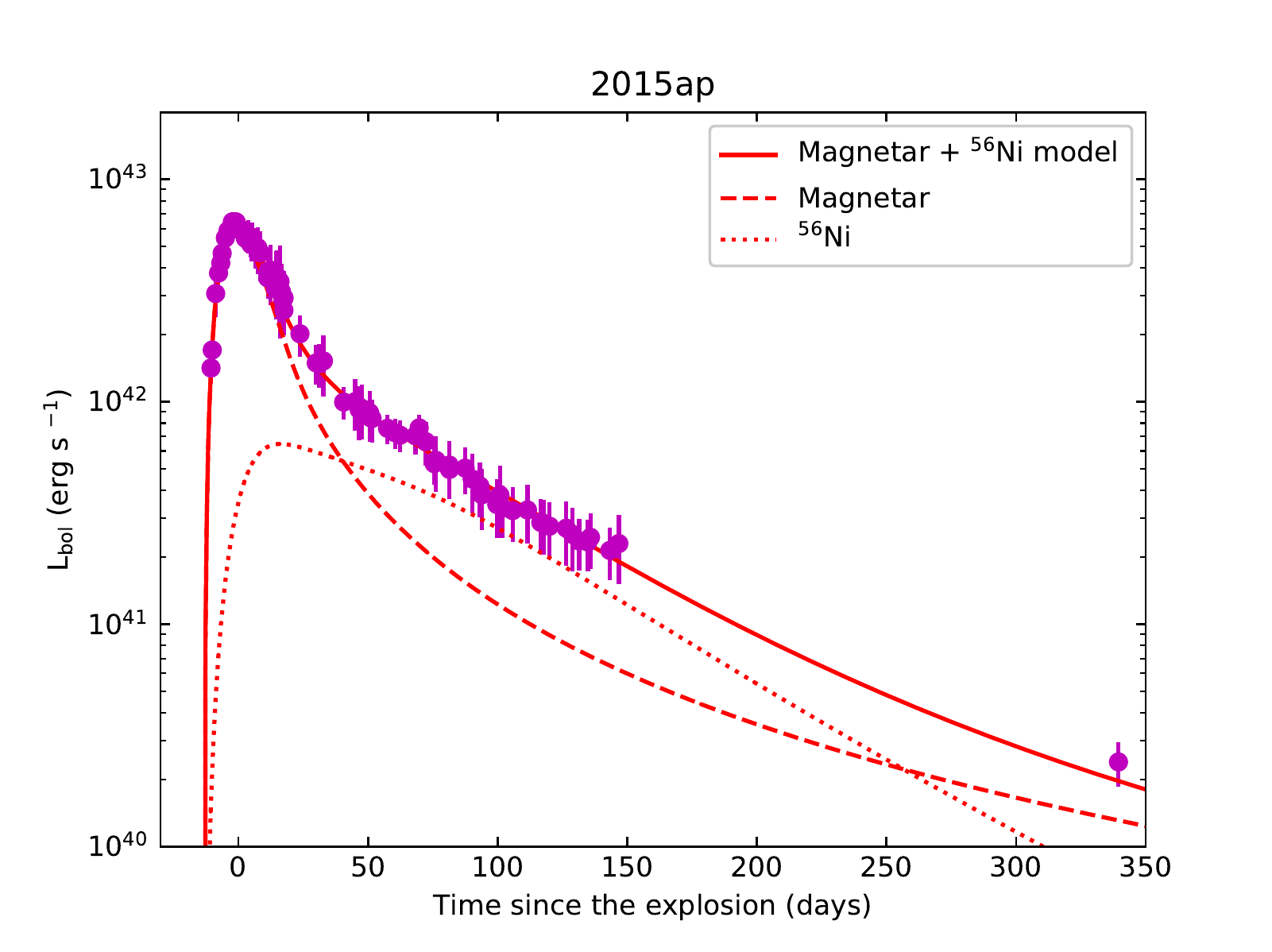}
		\includegraphics[width=0.45\textwidth,angle=0]{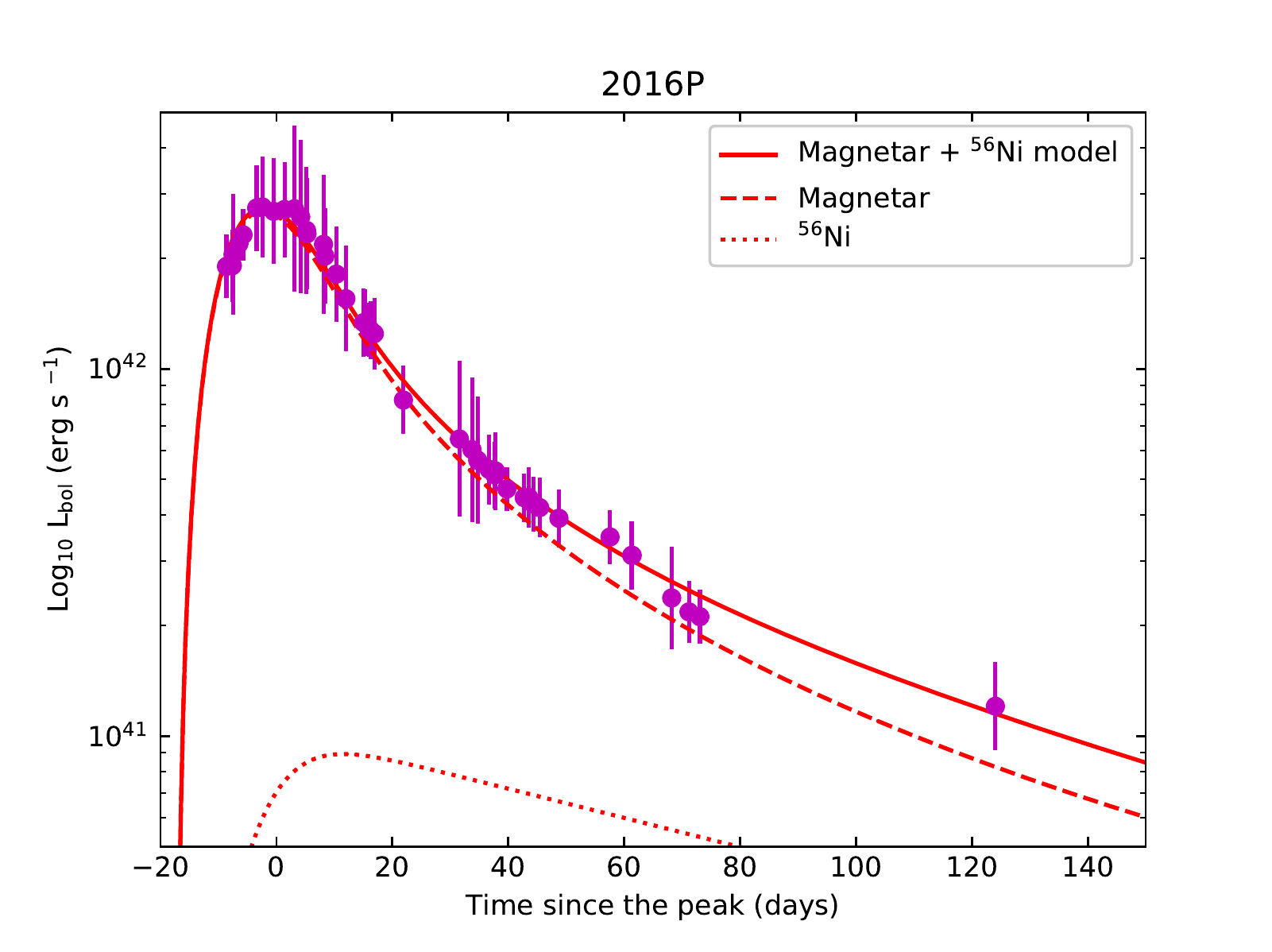}
	\end{center}
	\caption{The bolometric light curves of SNe 2015ap (the \textbf{left} panel) and 2016P (the \textbf{right} panel) reproduced by the Magnetar + $^{56}$Ni model (the \textbf{right} panel). The abscissa represents time since the explosion in rest frame.}
	\label{ni+mag-fits}
\end{figure*}
\begin{figure*}
	\begin{center}
		\includegraphics[width=0.8\textwidth,angle=0]{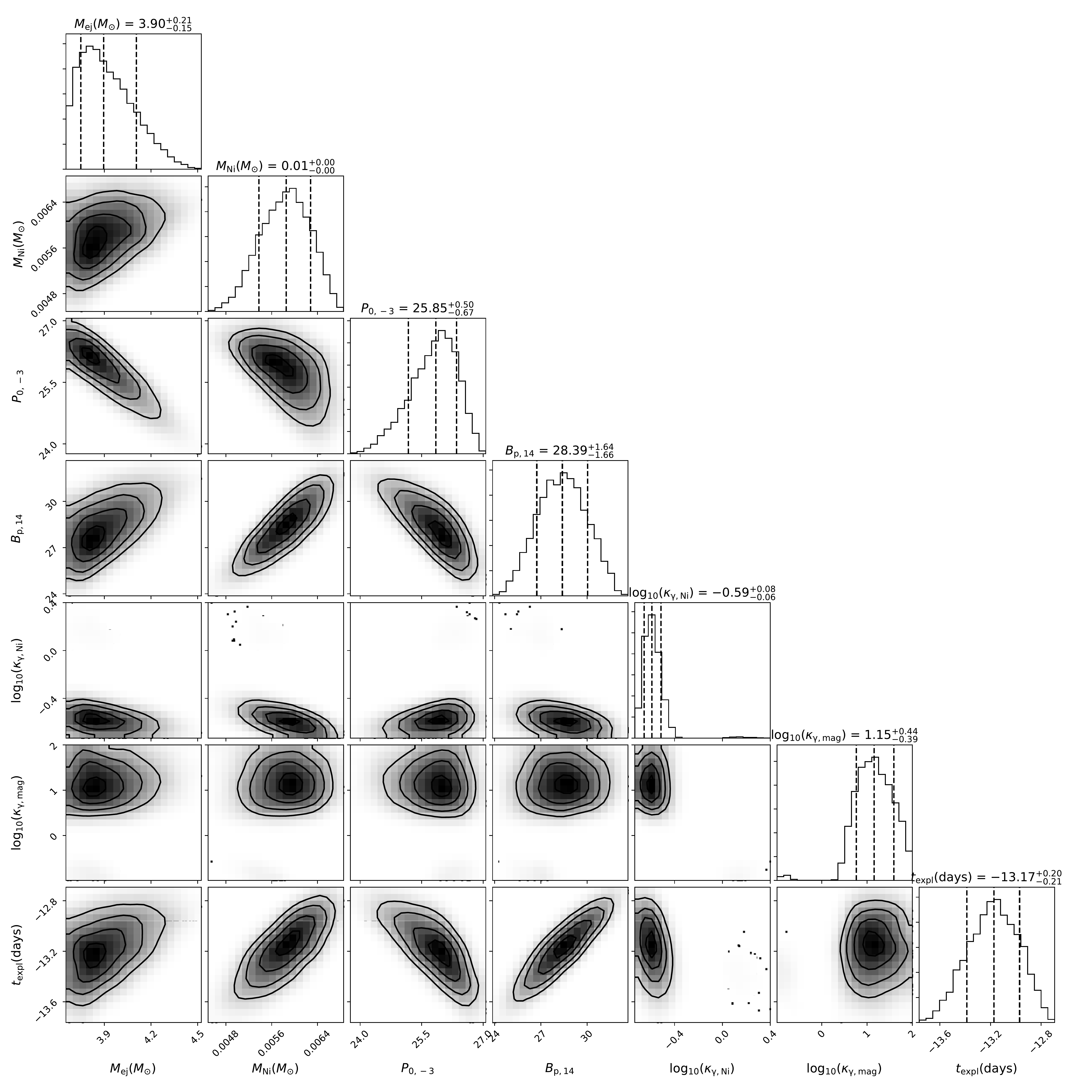}
	\end{center}
	\caption{Corner plot of SN 2015ap showing the parameters (M$_{\rm ej}$, M$_{\rm Ni}$, P$_{0}$, B$_{p}$, $\kappa_{\rm \gamma,Ni}$, $\kappa_{\rm \gamma,mag}$, t$_{\rm expl}$) of the $^{56}$Ni + magnetar model.}
	\label{corner2}
\end{figure*}
\begin{figure*}
	\begin{center}
		\includegraphics[width=0.8\textwidth,angle=0]{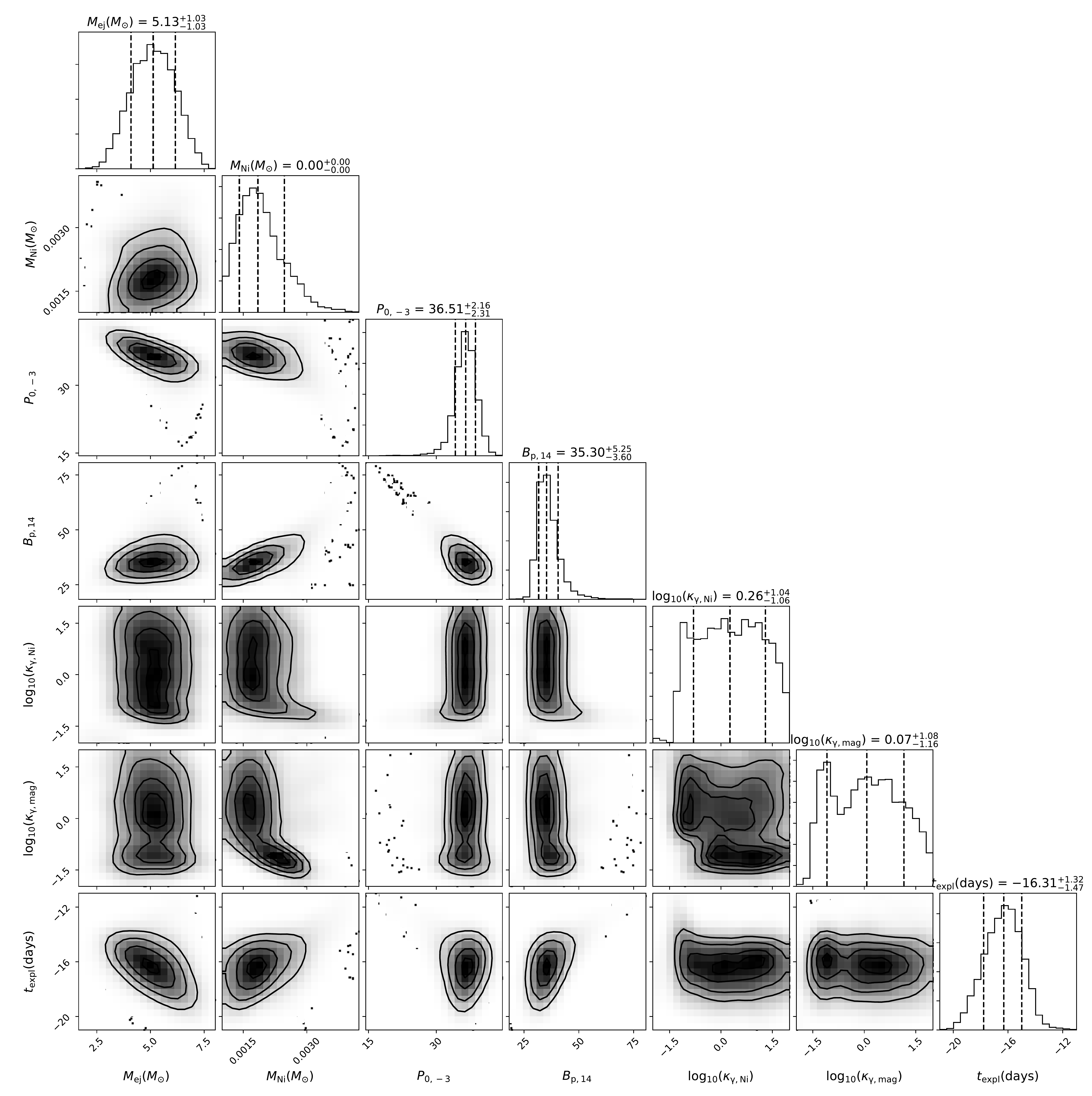}
	\end{center}
	\caption{Corner plot of SN 2016P showing the parameters (M$_{\rm ej}$, M$_{\rm Ni}$, P$_{0}$, B$_{p}$, $\kappa_{\rm \gamma,Ni}$, $\kappa_{\rm \gamma,mag}$, t$_{\rm expl}$) of the $^{56}$Ni + magnetar model.}
	\label{corner3}
\end{figure*}

The best-fit light curves are shown in Fig \ref{ni+mag-fits} and the parameters are tabulated in Table \ref{tab:para}. The associated corner plots showing confidence level of different parameters are also shown in Fig \ref{corner2} and Fig \ref{corner3} respectively. Even though the obtained $^{56}$Ni mass is low as compared to its kinetic energy, the $^{56}$Ni + magnetar model shows good fits to the overall light curves of both the SNe.
The uncertainties and the confidence of determination of the parameters as well as the degeneracy and correlation between parameters are shown in the corner plots. However, we remark that the values of ejected mass and the derived kinetic energy depend on the adopted value of opacity.

\section{Spectroscopic evolution}
\label{4}
Our spectroscopic observations of SN 2015ap started 14 days before $V$-maximum and continued upto 125 days post-maximum while the spectral coverage of SN 2016P is limited to 28 day after $V$-maximum. We explain in detail the spectroscopic features of these two events below.
\subsection{Spectral evolution of SNe 2015ap and 2016P during early phase}
\begin{figure}
	\begin{center}
		\includegraphics[width=0.5\textwidth]{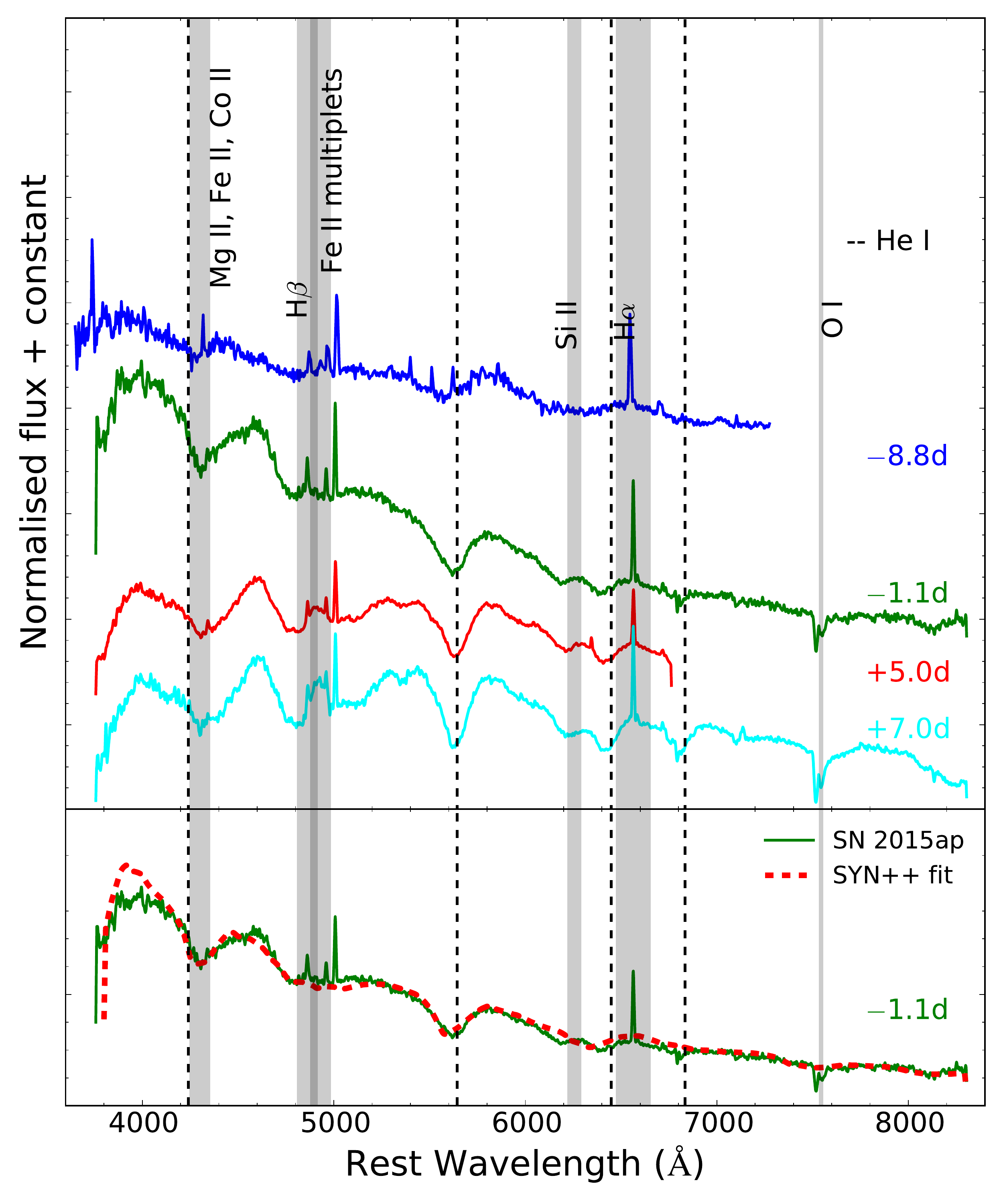}
	\end{center}
	\caption{Early time spectral evolution of SN 2015ap. Prominent He features are seen in early spectra. The bottom panel shows the $-$1.1 day spectrum plotted along with the best fit model generated by SYN++(dotted lines). Prominent lines of He I are marked with vertical dashed black lines.}
	\label{fig:specearly_2015ap}
\end{figure}

The pre-maximum spectral evolution of SN 2015ap is shown in Fig \ref{fig:specearly_2015ap}. The early time spectral sequence of SN 2015ap shows a blue continuum. Initial few spectra show distinct, blueshifted and broad P-cygni profiles which is indicative of very high velocities. However, we remark that since SN 2015ap exploded close to H II region, there is contamination of host C II and S II contamination lines. Moreover, around 5000 \AA~ we see significant contamination by H $\beta$ and [O III] galaxy lines. In the $-$1.1 day spectrum of SN 2015ap, He I 5876 \AA~ absorption feature expanding with a velocity of $\sim$ 15,500 km s$^{-1}$ is seen. Other prominent spectral features noticed in the early time spectra are Fe II between 4100 and 5000 \AA; Si II/H$\alpha$ at $\sim$ 6250 \AA~ and OI 7774 \AA. The He I 4471 \AA~ line is present but with a possible blend of Fe II complexes and Mg II 4481 \AA. Except 5015 \AA~ feature, all other He I lines at 4471, 5876, 6678 and 7065 \AA~ are prominent. The spectra taken at 5 day and 7 day of SN 2015ap in Fig \ref{fig:specearly_2015ap} show ``W" shape absorption feature around 4000 \AA. Similar kind of feature was observed in Type II SN 2005ap \citep{2007ApJ...668L..99Q}, Type Ib SN 2008D \citep{2009ApJ...702..226M}, SN 2009jf \citep{2011MNRAS.413.2583S} and Type IIb SN 2001ig \citep{2009PASP..121..689S} but at very early epochs (between $-$14 day to $-$10 day). While \cite{2008Sci...321.1185M} claim the origin of this feature due to Fe complexes, \cite{2009ApJ...702..226M,2009PASP..121..689S} consider this feature to be originated due to C III, N III and O III lines at high velocities. Since we see the feature in post-maximum phases, we consider the feature to be originated due to Fe complex with a blend of Mg II. The observed spectrum of SN 2015ap is compared with a synthetic spectrum generated using SYN++ \citep{2011PASP..123..237T} as shown in the bottom panel of Fig \ref{fig:specearly_2015ap}.
The synthetic spectrum with photospheric temperature of 13,000 K and photospheric velocity {\it v$_{\rm ph}$} = 16,000 km s$^{-1}$ reproduces the spectrum of SN 2015ap obtained at $-$1.1 day. The He I lines are moving with an expansion velocity of $\sim$ 15,500 km s$^{-1}$. \cite{2019MNRAS.485.1559P} also estimated the He I 5876 \AA~ line velocities between $-$ 10 days to 20 days post bolometric maximum which varied between 16,000 km s$^{-1}$ to 11,000 km s$^{-1}$ which is in agreement with our results. We performed a blackbody fit on the $-$ 1.1 day spectrum of SN 2015ap resulting in a temperature estimate of 12,000 $\pm$ 2,000 K. \cite{2019MNRAS.485.1559P} also estimated the temperature variations between $-$10 days to 40 days post bolometric maximum and found those to vary between 11,000 K to 5,000 K. Their estimated value of temperature are well within the errors of our estimates using blackbody fit.
\begin{figure}
	\begin{center}
		\includegraphics[width=0.5\textwidth]{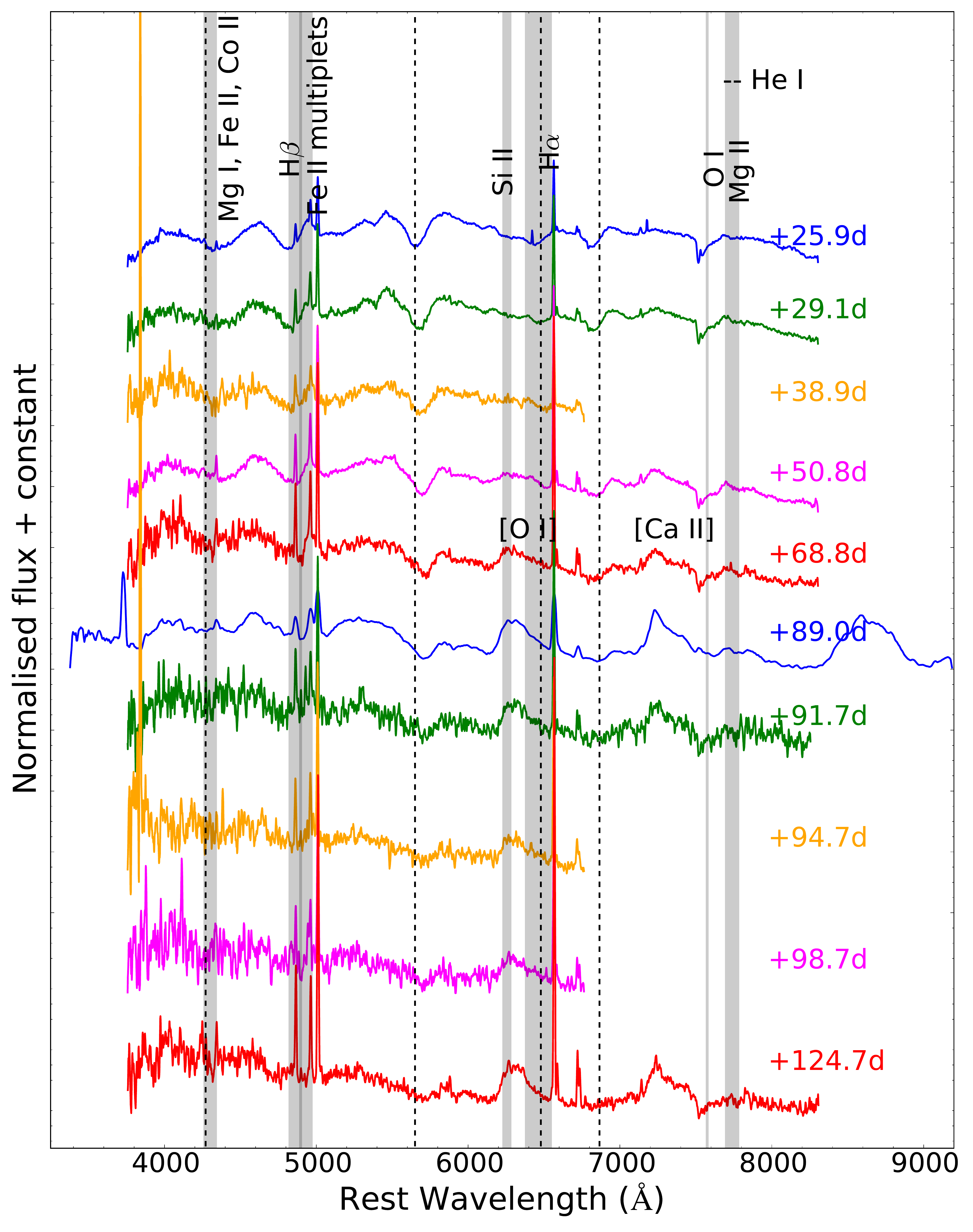}
	\end{center}
	\caption{Spectral evolution of SN 2015ap between 25 to 124 day post $V$-maximum. Post 70 days, the [O I] and [Ca II] features start dominating.}
	\label{fig:specmid_2015ap}
\end{figure}
Fig \ref{fig:specmid_2015ap} shows the evolution of SN 2015ap from 26 day to 124 day post maximum. We see that the broad He I absorption feature has decreased with time. Also, the blue continuum has decreased implying a decrease in photospheric temperature. All other spectral lines become prominent and well-developed. The He I line around 5876 \AA~ may be blended with Na I 5890, 5896 \AA. Post 90 day, we see distinct [O I] 6300, 6364 \AA~ doublet in the spectra of SN 2015ap which marks the onset of nebular phase. We notice prominent [O I] 6300, 6364 \AA~ and [Ca II] 7291, 7324 \AA~ doublet feature in Fig \ref{fig:specmid_2015ap} after this phase. Other identified lines in the spectra include Mg I] 4571 \AA~ feature, OI 7774 \AA~ blended with contributions from OI 8446 \AA. We see multipeaked [O I] in the nebular phase spectra of SN 2015ap which is indicative of a highly asymmetric ejecta configuration which will be discussed in detail in Section \ref{5}.
\begin{figure}
	\begin{center}
		\includegraphics[width=0.5\textwidth]{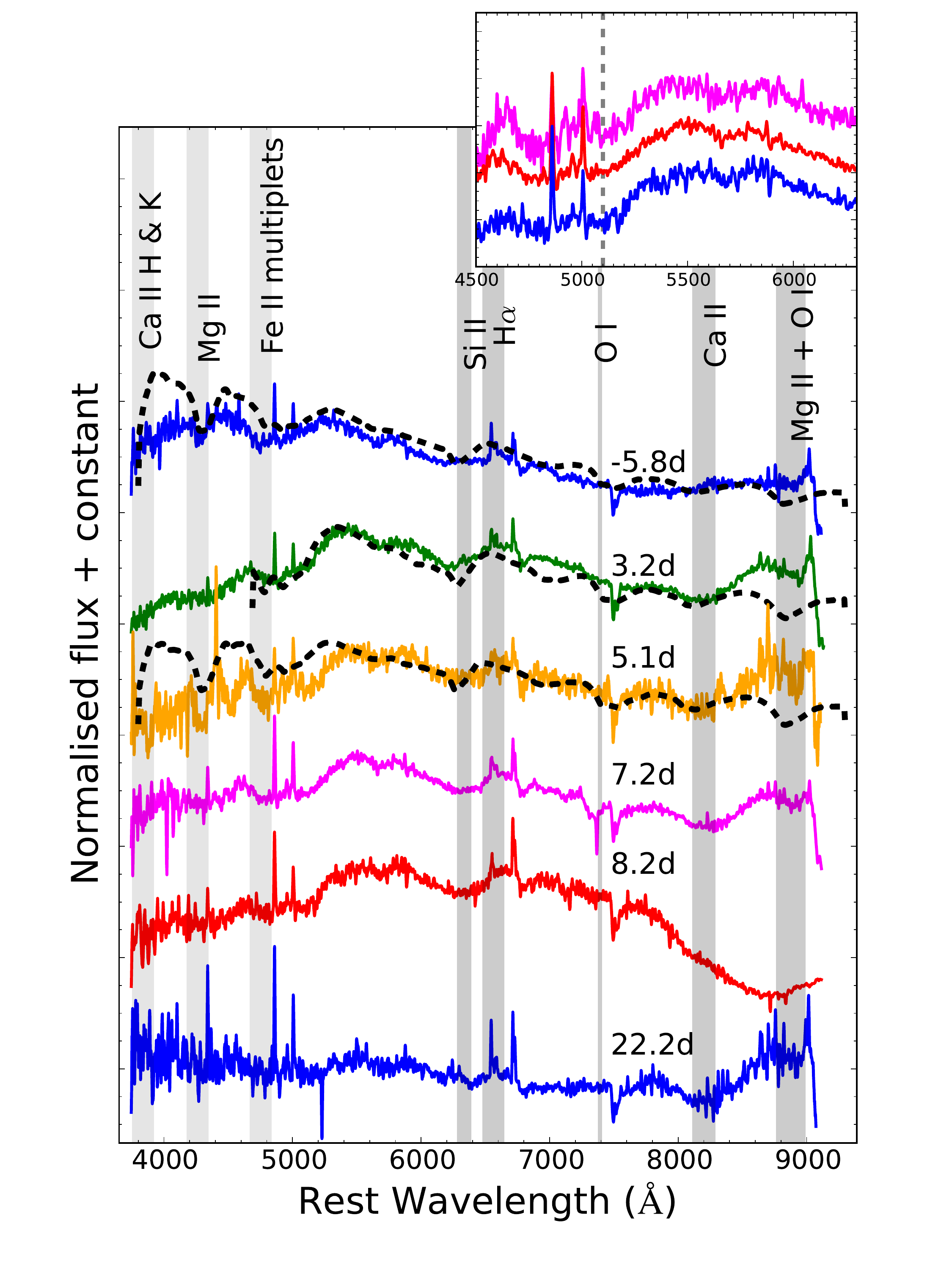}
	\end{center}
	\caption{Spectral evolution of SN 2016P. Dotted lines in the first three spectra show the best fit SYN++ model. The shaded region at 5100 \AA~ in the inset plot shows a shallow feature of unknown origin.}
	\label{fig:specearly_2016P}
\end{figure}

Fig \ref{fig:specearly_2016P} shows spectral sequence of SN 2016P covering a phase range of -5 to 22 days post $V$-maximum. The 0.1 day spectrum shows a blue continuum. Prominent lines of Mg II, multiplets of Fe II, Si II, O I and Ca II NIR features are seen in the spectrum. In the spectral sequence, we see two shallow absorption feature at 5100 \AA~ and 5500 \AA. The shallow absorption at 5150 \AA~ was noticed as an unidentified feature for Type Ic SN 2004aw \citep{2006MNRAS.371.1459T}. The broad absorption feature post 6000 \AA~ is believed to have several origins. The feature is thought to be associated with Si II 6355 \AA~   \citep{2003PASP..115.1220F,2008MNRAS.383.1485V,2009A&A...508..371H,2009ApJ...697..676S,2011ApJ...728...14P,2018MNRAS.473.3776K}. This absorption feature is also seen in SN 2007gr \citep{2008ApJ...673L.155V}, SN 2004aw \citep{2006MNRAS.371.1459T} and SN 2013ge \citep{2016ApJ...821...57D}. For SNe 2004aw and 2013ge, the feature is considered due to C II 6580 \AA. The absorption dip at 5500 \AA~ may result from a possible blending of Na I along with He I features \citep{2003A&A...408L..21P,2008MNRAS.383.1485V,2009ApJ...697..676S}.
To the first three spectra of SN 2016P, we fit SYN++ model. The dotted lines in Fig \ref{fig:specearly_2016P} shows the best-fit SYN++ model that is moving  with a photospheric velocity of {\it v$_{\rm ph}$} = 15,000 km s$^{-1}$ and temperature of 8000 K  which fits the first spectrum well. As we move to the third spectrum, {\it v$_{\rm ph}$} = 13,000 km s$^{-1}$ is achieved with temperatures reaching upto 7000 K. \cite{2019MNRAS.485.1559P} also show that at similar epochs, photospheric temperature of SN 2016P is 8,000 K. Prominent lines of Si II, Fe multiplets and Ca lines are also well reproduced. Si II lines show a velocity of 10,000 km s$^{-1}$ well in agreement with \cite{2019MNRAS.485.1559P}.
\subsection{Spectral comparison among SE-SNe}
\begin{figure}
	\begin{center}
		\includegraphics[width=0.5\textwidth]{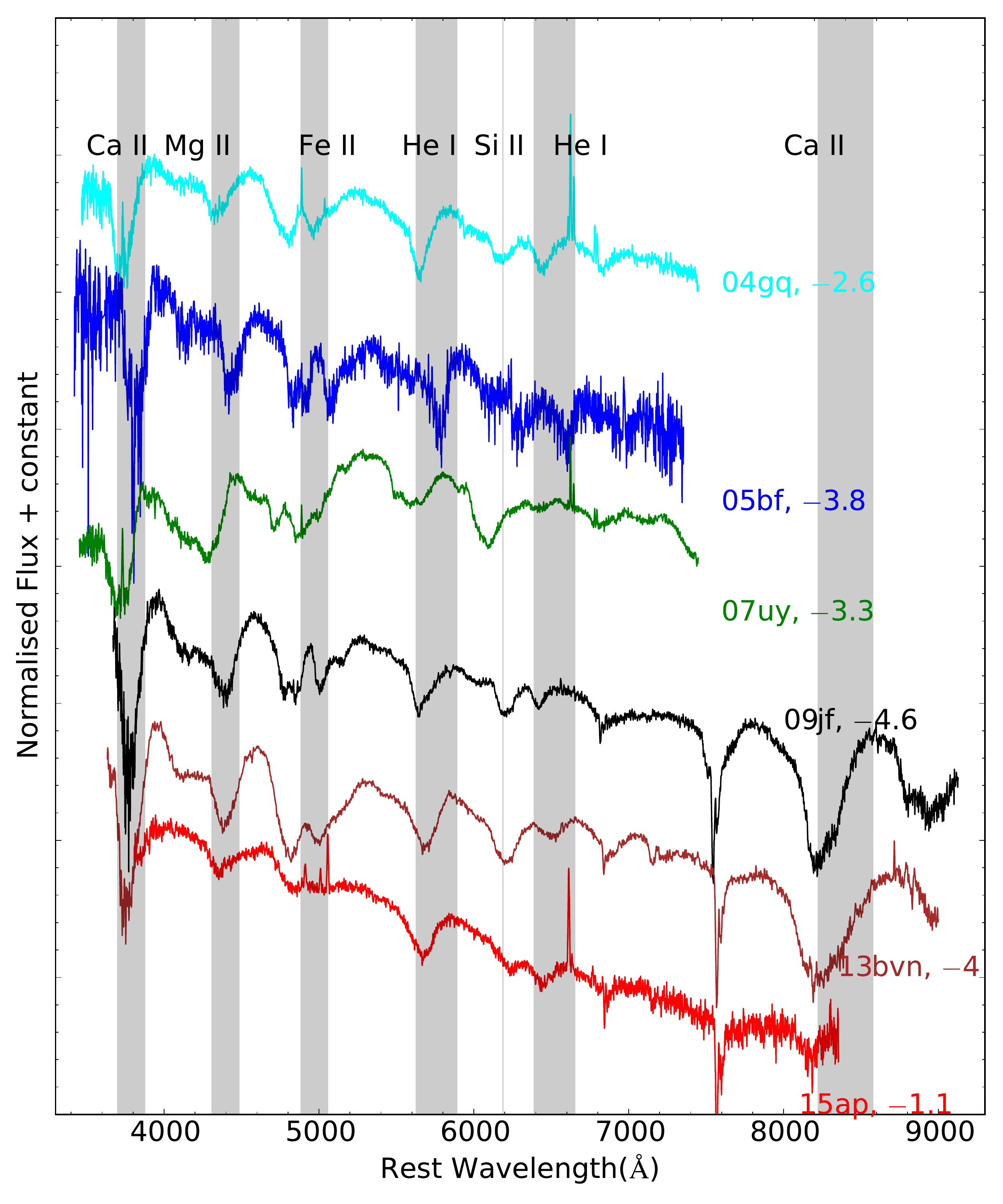}
	\end{center}
	\caption{Spectral comparison of SN 2015ap along with other Type Ib SNe. SN 2015ap shows a distinct blue continuum.}
	\label{fig:compearly_2015ap}
\end{figure}
The comparison of the pre-maximum spectrum of SN 2015ap is shown in Fig \ref{fig:compearly_2015ap}. The $-$1.1 d spectrum of SN 2015ap shows a distinct blue continuum as compared to all other SNe indicating higher temperatures. This is also in agreement with the results reproduced by SYN++ fitting. He I absorption lines are well established in all the spectra.

\begin{figure}
	\begin{center}
		\includegraphics[width=0.5\textwidth]{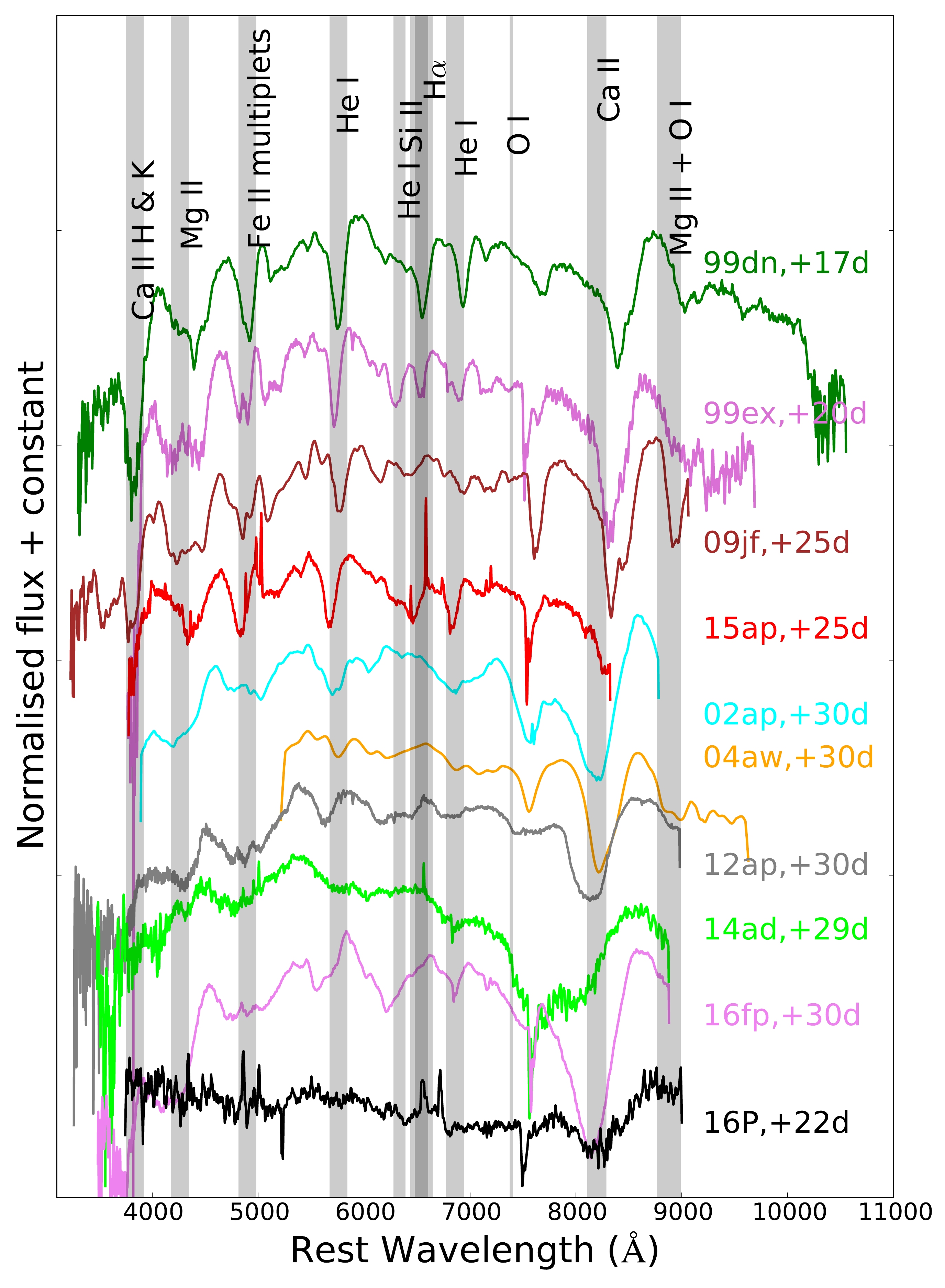}
	\end{center}
	\caption{Spectral comparison of SNe 2015ap and 2016P at $\sim$ 25 day post $V$-maximum with other SNe. The overall spectrum of SNe 2015ap and 2016P shows resemblance with SNe 2009jf and 2012ap respectively.}
	\label{fig:comp_2016}
\end{figure}

We compare the $\sim$ 25 day spectra of SNe 2015ap and 2016P with a number of Ib, Ic, Ic-BL and transitional Ic SNe in Fig \ref{fig:comp_2016}. The spectrum of SN 2015ap bears a close resemblance with SN 2009jf. Prominent He I 5876 \AA~ features are seen in the whole sample of Type Ib SN. The He I features in SN 2015ap are very similar to SN 2009jf. The overall spectrum of SN 2016P (at 22 day) is quite similar to SN 2012ap. Prominent Mg II features at other wavelengths along with Mg II + O I at 9000 \AA~ is well-developed as compared to all other SNe of the comparison sample. The absorption dip close to 6200 \AA~ is also found to be similar to that of SN 2012ap. The overall spectra of SN 2016P shows lines having width in between Type Ic and Ic-BL SNe.

\begin{figure}
	\begin{center}
		\includegraphics[width=0.5\textwidth]{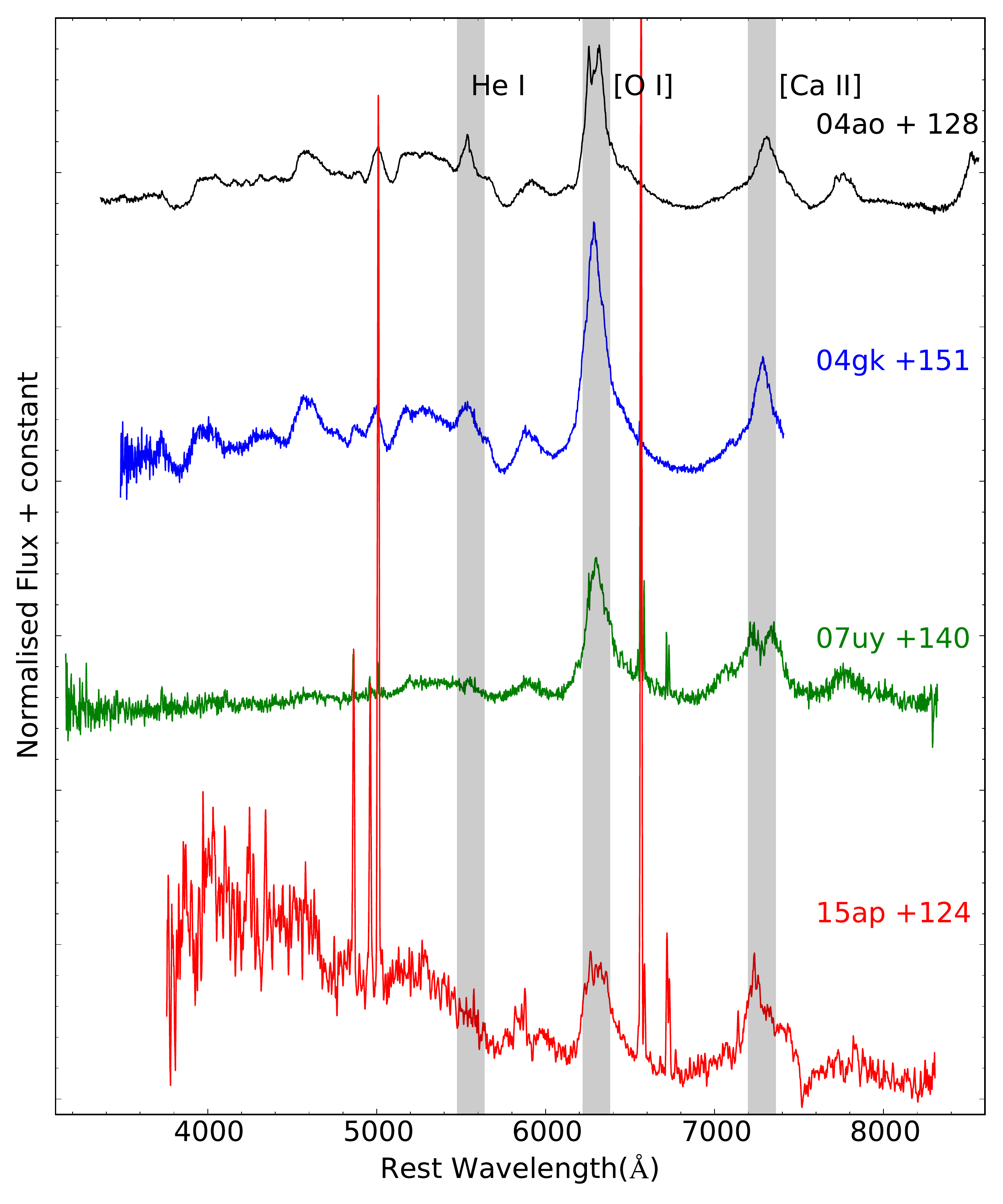}
	\end{center}
	\caption{Nebular phase comparison of SN 2015ap shows less prominent emission peaks than other SNe. SN 2015ap shows distinct multi-peaked profiles of [O I] different from all other SNe.}
	\label{fig:complate_2015ap}
\end{figure}
Fig. \ref{fig:complate_2015ap} shows the nebular phase comparison spectra of SN 2015ap along with other Type Ib SNe. SN 2015ap shows an asymmetric emission profile of [O I] as compared to other SNe. Also, the [Ca II] doublet is broad and dispersed than others. This has several implications on the physical scenario which we will explain in detail in Section \ref{5}.

\section{Asymmetry in the nebular phase emission of SN 2015ap}
\label{5}

As the nebular phase probes deeper into the ejecta, several studies on the ejecta geometry and associated asymmetries have been done in the past \citep{2005Sci...308.1284M,2008Sci...319.1220M,2008ApJ...687L...9M,2009MNRAS.397..677T,2010ApJ...709.1343M}. The most-prominent lines in nebular phase to study ejecta geometry are [O I] 6300, 6364 \AA~ doublet, [Ca II] 7291, 7324 \AA~ doublet and the Mg I] 4571 \AA~ lines. However, [O I] is best to probe asymmetry as [Ca II] lines may be contaminated with [O II] 7320, 7330 \AA~ \citep{1999AJ....117..725F} lines and Mg I] line is contaminated by Fe lines. Also, oxygen is the most abundant element in the SN ejecta and forms in a wavelength region where sensitivity of most of the spectrographs is maximum.
The ratio of [O I] 6300/6364 \AA~ lines changes from 1:1 to 3:1 for H-rich SNe when the density limit approaches 10$^{10}$ cm$^{-3}$ \citep{1991ApJ...372..531L,1991BAAS...23R.881P,1992SvAL...18..239C}. While \cite{2009MNRAS.397..677T} suggested that during nebular phase, the ratio is always 3:1 for Type Ib SNe but \cite{2010ApJ...709.1343M} argued that the ratio may be close to 1:1 in order to explain the dominance of double peaks in Type Ib rather than Type Ic. \cite{2006ApJ...645.1331M} used a ratio of 3:1 for their nucleosynthesis models which was observationally confirmed by \cite{2009A&A...508..371H} for SN 2007gr. \cite{2010MNRAS.409.1441M} showed that high velocity H$\alpha$ absorption feature can be superimposed on the oxygen emission profiles giving rise to double peaks.

\begin{figure*}
	\begin{center}
		\includegraphics[width=\textwidth]{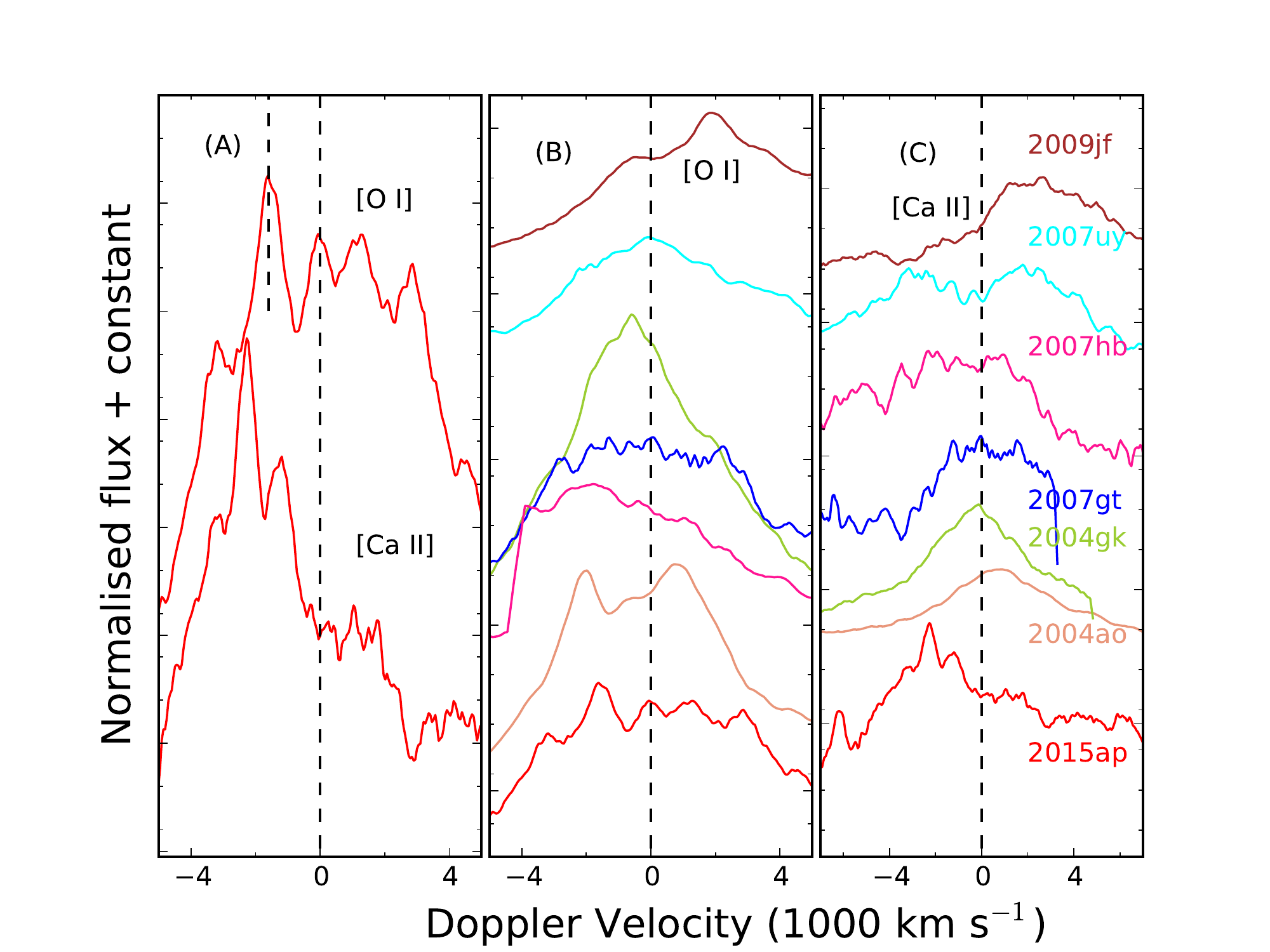}
	\end{center}
	\caption{Panel (A) of the Fig represents the line profiles of [O I] 6300, 6364 \AA~ and [Ca II] 7291, 7324 \AA. Panel (B) and (C) shows the comparison of the [O I] and [Ca II] profile of SNe 2015ap with other Type Ib/c SNe sample. The black dashed lines in Fig 18 (A) represents the zero velocity line corresponding to 6300 \AA~ and left line represents the real peak at -1600 km sec$^{-1}$. In panel (B), the black line represents the velocity corresponding to 6300 \AA~ and panel (C) represents the velocity corresponding to 7291 \AA~ line.}
	\label{fig:ca_o_evol}
\end{figure*}

\begin{figure*}
	\begin{center}
		\includegraphics[width=\textwidth]{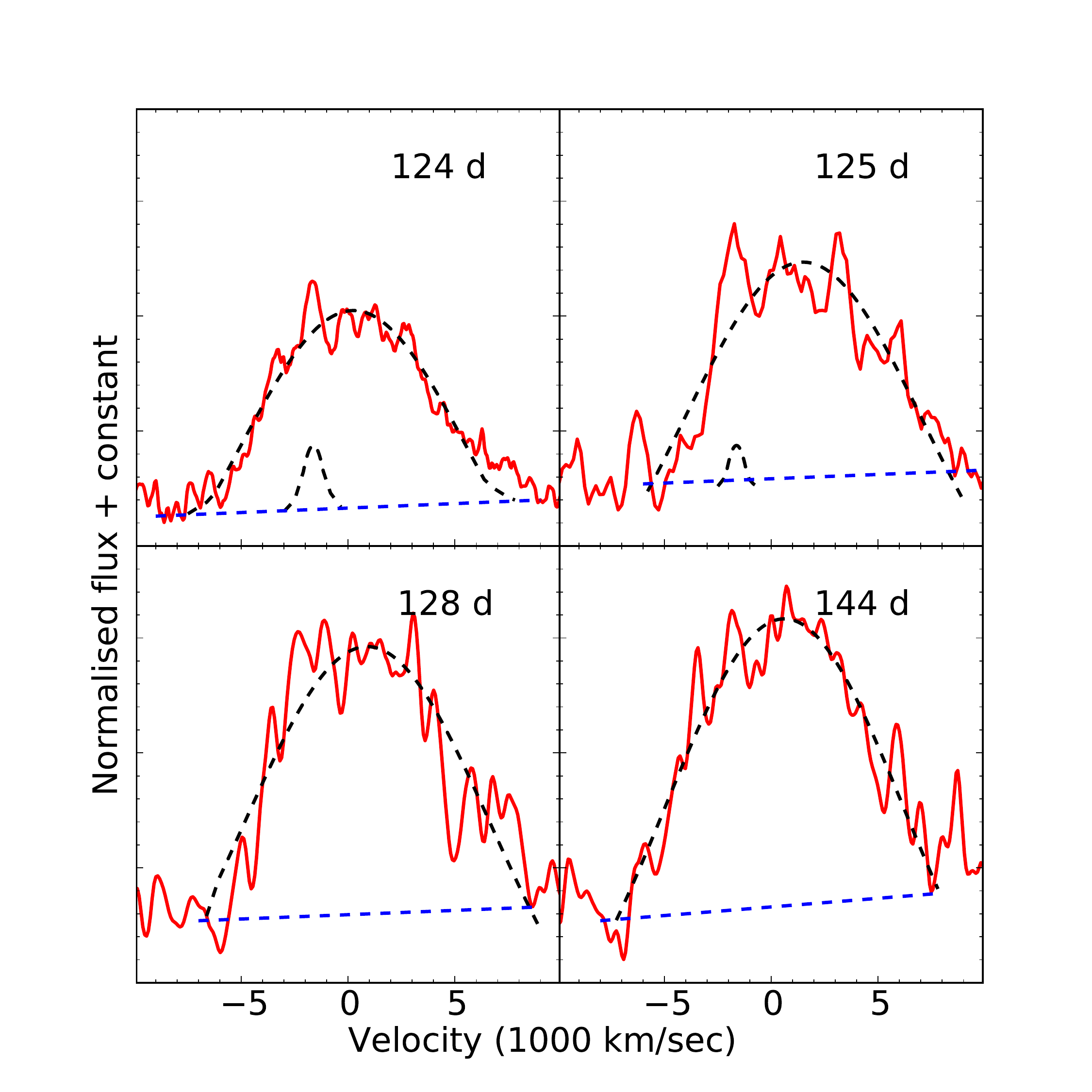}
	\end{center}
	\caption{Oxygen profile of SN 2015ap fitted with Gaussian profiles. Top two panels shows two peaks while bottom two panels shows an overall asymmetric structure.}
	\label{fig:o_evol}
\end{figure*}

SN 2015ap has a complex oxygen profile with multiple peaks/asymmetric shifted from 6300 \AA~ as shown in Fig \ref{fig:ca_o_evol}(A). We do not detect any Hydrogen features in SN 2015ap, so the high velocity feature as the reason for multi-peaked profile is ruled out. The spectrum has been smoothed using a boxcar factor of 9 to reduce the noise.   The SNR of all the peaks was checked in the smoothed spectrum of 124 d and compared with the background. The overall spectrum has a SNR of about 4. We checked the SNR of the individual peaks and found that the peak at -1600 km/sec is of course real and is four times higher than the background, thus validating that the peak is real and hence confirming existence of an asymmetric profile. The other peaks are two times higher than the background. As we obtain a blueshifted peak, we interpret this as per \cite{2009MNRAS.397..677T} which explains blueshift being originated due to either suppression of the redshifted part of the spectrum as dust forms or residual opacity as ejecta cools. These kind of profiles have been seen previously in SNe 1990B, 1990aa, 1999dn, 2006F and 2009jf \citep{2009MNRAS.397..677T}. \cite{2009MNRAS.397..677T} explained these asymmetric profiles that are produced by additional components of arbitrary widths and shifts with respect to main component. This indicates clumping of the SN ejecta on a large scale. It may also be due to a single massive blob and a unipolar jet.  However, to check the evolution of line profiles, we check this in the three additional spectra taken from WiseREP public archive\footnote{https://wiserep.weizmann.ac.il} and repeat the fitting procedure as discussed above. We find that the spectra of 124 d and 125 d are well fit with two Gaussians but the spectra of 128 d and 144 d are well fit with a single Gaussian (see Fig \ref{fig:o_evol}).  This indicates a transition in the [O I] line profile, but an overall clumpiness and asymmetric ejecta configuration is seen in all the profiles.

Fig \ref{fig:ca_o_evol}(A) also displays the Ca II profile. We see a blueshifted prominent peak in the doublet. This has a similar origin as that of [O I] profile  \citep{2009MNRAS.397..677T}. The overall variation is however smooth as compared to the [O I] emission. Clumps of Ca do form, but they do not contribute significantly to emission. These clumps of Ca intercept $\gamma$-ray radiation. However, as the mass fraction is small, the amount of radioactive luminosity and temperature achieved is not sufficient for the emission \citep{1993ApJ...419..824L, 2000AJ....120.1487M}. The pre-existing envelope of the star is heated generally by hydrogen and helium that do not radiate efficiently, allowing Ca that was thoroughly mixed into the atmosphere, most likely present from the time of the progenitor formation, to reach the temperatures required to produce [Ca II] emission indicating a primordial origin.

 Fig \ref{fig:ca_o_evol}(B) and  Fig \ref{fig:ca_o_evol}(C) compares the spectrum of SN 2015ap with a number of Type Ib/c SNe (phase range - 110 to 150 day) that are taken from WiseREP public archive. We see that the [O I] profile of SN 2015ap bears a resemblance with SNe 2007gt and 2009jf while [Ca II] profile bears a close resemblance with SN 2007hb. The [O I] profile of SN 2015ap shows a clumpy profile as compared to the double peaked and single peaked profile of most of the other SNe. This may be due to the different geometrical orientation of the [O I] in different SN ejecta. If we look into the [Ca II] profiles, the profiles of all the Type Ib SNe have clumpiness less than the [O I] profiles. This may be due to the fact that the amount of primordial Calcium is more than the [Ca II] that are generated during the nucleosynthesis process going on in the SN evolution.

\subsection{Estimates of the O mass and [O I] / [Ca II] ratio}

The line strengths in the nebular phase spectra provide key information on the progenitor mass. \cite{1986ApJ...308..685U} derived a relation, showing the minimum mass of O that can be estimated in the high density  (N$_{e}$ $\geq$ 10$^{6}$ cm$^{-3}$) limit \citep{1989AJ.....98..577S,2004A&A...426..963E}. This is given by
\vspace{-0.1cm}
\begin{equation}
M_{O} = 10^{8} \times D^{2} \times F([O\,I]) \times exp^{(2.28/T_{4})},
\end{equation}
where $M_{O}$ is the mass of the neutral O in M$_{\odot}$ units, $D$ is distance to the galaxy in Mpc, $F([O I])$ is the total flux of the [O I] 6300, 6364 \AA~ feature in erg s$^{-1}$ cm$^{-2}$, and $T_{4}$ is the temperature of the O-emitting region in units of 10$^{4}$ K. Ideally, the ratio of [O I] 5577 \AA~to the [O I] 6300, 6364 \AA~ feature should be considered. \cite{1989S&T....78..491O} showed that the ratio of [O I] 5577 \AA~ to [O I ] 6300 is proportional to the electron density and the temperature of the emitting regions. Thus, the uncertainties associated in the equation are large which involves the errors due to the flux, temperature and indirectly the density. The temperatures can typically vary between (0.4 -- 1) x 10$^{4}$ K. We find that the [O I] 5577 \AA~ line is very faint, and the limit of the flux ratio is $\leq$ 0.1. Using the observed flux of 3.0 x 10$^{-14}$ erg s$^{-1}$ cm$^{-2}$ of the [O I] 6300, 6364 \AA~ doublet from the 2016 January 22 spectrum, and adopting T$_{4}$ = 0.4 K, we estimate $M_{O}$ = 0.90 M$_{\odot}$.  A very weak OI 7774 \AA~ emission feature is seen in the nebular spectra. Since the O I 7774 line is expected to result from recombination of ionized O \citep{1986ApJ...302L..59B}, the O mass required to produce the [O I] doublet and the O I 7774 \AA~ line is higher than the mass required to produce the doublet alone \citep{2010MNRAS.408...87M}. Therefore, 0.90 M$_{\odot}$ can be considered as a lower limit of total oxygen mass ejected in the explosion. [O I] layer is mostly produced by hydrostatic burning of oxygen. \cite{1996ApJ...460..408T} made explosive nucleosynthesis calculations and predicted major nucleosynthesis yields for the progenitor mass of 13 $-$ 25 M$_{\odot}$. For progenitor masses of 13, 15, 20 and 25 M$_{\odot}$, \cite{1996ApJ...460..408T} showed that the corresponding O masses would be 0.22, 0.43, 1.48 and 3.0 M$_{\odot}$, respectively. Also, \cite{1996ApJ...460..408T} estimated He core masses of 3.3, 4 and 8 M$_{\odot}$, corresponding to progenitors of 13, 15 and 25 M$_{\odot}$, respectively. Thus, using this model and corresponding approximations we conclude that the progenitor of SN 2015ap is most likely a star between 15 $-$ 20 M$_{\odot}$ and He core mass between 4 $-$ 8 M$_{\odot}$.

\cite{2015A&A...579A..95K} estimated the value of [O I]/[Ca II] ratio for a group of core-collapse SNe and found that this ratio is highly dependent on temperature, density and progenitor mass. \cite{2015A&A...579A..95K} considered a demarcation of SNe progenitors as binary and single if this ratio was lower or higher than 1.5 respectively. However, this ratio is close to 0.7 for Type IIP SNe. In the case of SN 2015ap the [O I]/[Ca II] ratio is $\sim$ 0.71 indicating that the progenitor is a low mass star in a binary system.

\subsection{Progenitor mass from oxygen line luminosities}
\begin{figure}
	\begin{center}
		\includegraphics[scale=0.45]{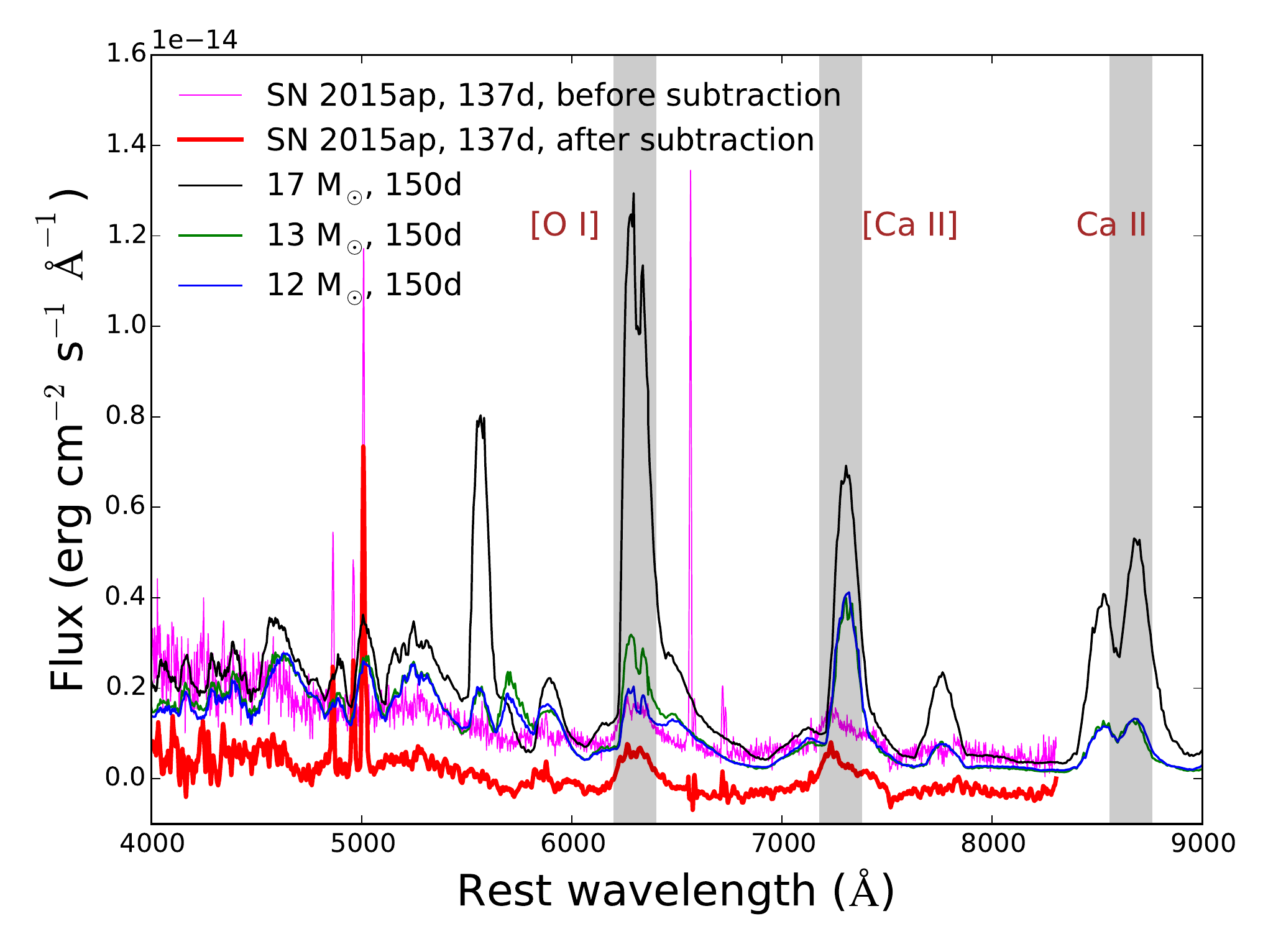}
	\end{center}
	\caption{The 137 day since explosion spectrum of SN 2015ap plotted along with the 12, 13 and 17 M$_{\odot}$ models of Jerkstand et al. 2015 at 150 day. The 12 M$_{\odot}$ model best matches our observed spectrum of SN 2015ap.}
	\label{fig:Jerkstrand}
\end{figure}

The observed strength of the nebular lines can serve as an important diagnostic of the progenitor mass. \cite{2012A&A...546A..28J,2014MNRAS.439.3694J} generated nucleosynthesis models for progenitor masses of 12, 15, 19 and 25 M$_{\odot}$, assuming $^{56}$Ni mass to be 0.062 M$_{\odot}$ and distance of 5.5 Mpc for a group of Type IIP SN. \cite{2015A&A...573A..12J} extended this for a group of SE-SNe by implementing minor modifications in the code. They generated three models of 12, 13 and 17 M$_{\odot}$ for the stripped envelope group with $^{56}$Ni mass = 0.079 M$_{\odot}$ and distance of 7.9 Mpc. In Fig \ref{fig:Jerkstrand}, we compare the 137 day post explosion (or 124 day since $V$-maximum) spectrum of SN 2015ap with the 150 day and 12, 13 and 17 M$_{\odot}$ model generated by \cite{2015A&A...573A..12J} for Type IIb/Ib SNe. It is to be noted that the model assumes elemental abundances which are higher than the estimates of  for lower progenitor mass.

We modelled the spectra of SN 2015ap adopting a ratio of (0.26/0.079) to correct $^{56}$Ni mass and rescale the model flux (150 d) to the distance and phase of the (137 day) nebular spectrum of SN 2015ap. The SN spectrum is de-reddened and de-redshifted to avoid any flux loss before comparison. We see that the overall spectrum matches better with a 12 M$_{\odot}$ progenitor star. We then, subtracted the host galaxy spectrum taken from SDSS DR14 of SN 2015ap from the SN spectrum. Fig \ref{fig:Jerkstrand} shows both the subtracted and unsubtracted spectrum of SN 2015ap. The galaxy lines of H and S II are removed in this spectrum and also there is an overall flux decrement, implying a progenitor of mass less than 12 M$_{\odot}$. It is to remark, however, that these models assumes an over abundance of elements as compared with \cite{1996ApJ...460..408T}, so a lower progenitor mass would indicate an over estimated peak. Also, due to the high density envelope used in the model, scattering of other lines may affect the flux estimations. The [O I] feature matches well with a 12 M$_{\odot}$ progenitor star for the unsubtracted spectrum. Our obtained value of {\it L$_{norm}$} = 0.015  which is a good tracer of {\it L$_{[OI]}$} \citep{2015A&A...573A..12J} matches well with the value estimated by \cite{2019MNRAS.485.1559P}. Comparing the line luminosities of [O I] and the fitted nebular spectral modeling indicates that the progenitor of SN 2015ap is in the range of 12 $-$ 20 M$_{\odot}$ which is most likely in a binary association.
\subsection{Host galaxy metallicity}
Metallicity is a key parameter driving mass-loss rates in single stars and influences the relative number of Type Ib/c SNe. It's also important for binary models \citep{2008MNRAS.384.1109E} and affects the ratios of Type Ib/c SNe. In order to estimate the host galaxy metallicity, we used the Sloan Digital Sky Survey (SDSS) spectrum of IC 1776; the host galaxy of SN 2015ap. The DR12 SDSS image was taken on 2010 November 13\footnote{https://dr12.sdss.org/spectrumDetail?mjd=55514\&fiber=999
	\&plateid=4402}. Narrow emission lines of [O III] 5007 \AA~ and the [N II] 6583 \AA~ lines are clearly visible in the host galaxy spectrum. Several diagnostics are used to estimate the host galaxy metallicity \citep{1991ApJ...380..140M, 2002ApJS..142...35K, 2004MNRAS.348L..59P,2005ApJ...631..231P}. We used the O3N2 index calibration given by \cite{2004MNRAS.348L..59P} to estimate the host galaxy metallicity. The value of 12 + log(O/H) is found to be 8.8 $\pm$ 0.6 dex. For SN 2016P, the SDSS DR12 image of the galaxy NGC 5374 was taken on 2007 March 16\footnote{https://dr12.sdss.org/spectrumDetail?mjd=54176\&fiber=301
	\&plateid=1808}. In this spectrum also, narrow emission lines of [O III] 5007 \AA~ and the [N II] 6583 \AA~ were seen. The value of 12 + log(O/H) for NGC 5374 is found to be 8.7 $\pm$ 0.5 dex.
Several authors have estimated the solar metallicity values to be 8.69 $\pm$ 0.05 dex, 8.69 dex and 8.76 $\pm$ 0.07 dex,  \citep{2001ApJ...558..830A,2009ARA&A..47..481A,2011SoPh..268..255C}
The obtained metallicity value for IC 1776 is slightly super-solar while NGC 5374 has nearly solar metallicity. However, the metallicity is close to solar for both the galaxies within errors. Wolf-Rayet progenitor could be one of the favourable scenario as the number of WR stars increases with increasing metallicity due to the strong metallicity dependence on stellar winds \citep{2005A&A...442..587V,2010A&A...516A.104L}. The progenitors could be in the form of either single WR stars or in binary association. \cite{2016A&A...591A..48G} estimated the metallicity of a group of Type II and Ib/c SNe and found an average metallicity of 8.50 dex with an error of 0.02 for all SNe. Their results support a binary progenitor for SNe Ib and at least a fraction of SNe Ic to be single massive stars that have stripped their outer layers by metallicity driven winds.

\section{Summary}
\label{6}
In this paper, we present the temporal and spectral evolution of a Type Ib SN 2015ap and a Type Ic SN 2016P. The early decline rates of SNe 2015ap and 2016P are consistent with other Type Ib/c SNe while the late time light curves of both the SNe shows steepening. The colours of SN 2015ap are similar to most Type Ib comparison sample while the colour evolution of SN 2016P is redder than most of the members. The absolute magnitude (M$_{V}$ = $-$18.04 $\pm$ 0.19 mag) of SN 2015ap indicates that it lies at the brighter end of Type Ib SN among the comparison sample while SN 2016P lies in the middle (M$_{V}$ = $-$17.53 $\pm$ 0.14 mag) among Type Ic sample. The bolometric light curve modelling of SNe 2015ap and 2016P indicates that the light curves can be best fit by a $^{56}$Ni + magnetar model with $^{56}$Ni masses of 0.01 M$_{\odot}$ and 0.002 M$_{\odot}$, ejecta masses of 3.75 M$_{\odot}$ and 4.66 M$_{\odot}$, spin period P$_{0}$ of 25.8 ms and 36.5 ms and magnetic field B$_{p}$ of 28.39 $\times$ 10$^{14}$ Gauss and 35.3 $\times$ 10$^{14}$ Gauss respectively. We suspected an additional energy source from a newly born magnetar because of the inability to fit the late time light curves and high luminosity. The narrow light curve and short rise time for SN 2015ap implied that the high ejecta mass is most likely a resultant of terminal fallback.

Early time spectral sequence of SN 2015ap shows prominent He I absorption lines with a blue continuum and high photospheric temperature. SYN++ modelling of the early time spectra of SN 2015ap indicates a photospheric temperature of 13,000 K and a photospheric velocity of 16,000 km s$^{1}$. Initial spectral sequence shows a ``W" like feature arising due to Fe complexes. SN 2016P shows prominent lines of Ca II H $\&$ K, Mg II, Fe II and Ca II NIR features. The broad absorption blueward of 6000 \AA~ mostly arises due to a blend of Si II and C II. SYN++ modelling gives an initial photospheric velocity {\it v$_{\rm ph}$} = 15,000 km s$^{-1}$ and temperature of 8000 K.

The [O I] profile of SN 2015ap shows an asymmetric profile which is indicative of large scale clumping or a unipolar jet. The [O I] and [Ca II] doublet shows blueshifted peak originating from residual opacity or dust formation. [Ca II] envelope has an origin of pre-SN envelope. The estimated O mass is found to be 0.90 M$_{\odot}$. The [O I]/[Ca II] ratio and the Jerkstrand modelling indicates a progenitor of mass between 12 $-$ 20 M$_{\odot}$ and most likely in binary association \citep{2011MNRAS.416..817S}.

\section*{Acknowledgments}
We sincerely thank the referee for his/her valuable comments and input which has improved the presentation of the paper. We thank the observing staff and observing assistants at 1.04 m ST, 1.30 m DFOT, 1.82 m EKAR Asiago Telescope and 2.00 m HCT for their support during observations. We thank Andrea Pastorello for co-ordinating the observation of SN 2015ap with Asiago Telescope. We acknowledge Wiezmann Interactive Supernova data REPository http://wiserep.weizmann.ac.il (WISeREP).  BK acknowledges the Science and Engineering Research Board (SERB) under the Department of Science \& Technology, Govt. of India, for financial assistance in the
form of National Post-Doctoral Fellowship (Ref. no. PDF/2016/001563). LT is partially supported by the  ``PRIN-INAF 2017" with the project ``Towards the SKA and CTA era: discovery, localization, and physics of transient objects''. The work made use of the {\it Swift} Optical/Ultraviolet Supernova Archive (SOUSA). SOUSA is supported by NASA's Astrophysics Data Analysis Program through grant NNX13AF35G. SBP and KM acknowledges BRICS grant DST/IMRCD/BRICS/Pilotcall/ProFCheap/2017(G) for the present work. N.E.R. acknowledges support from the Spanish MICINN grant
ESP2017-82674-R and FEDER funds.









\appendix

\section{Photometry}
In addition to the ground based telescopes, SN 2015ap was observed with {\it Swift} UVOT  (\citealp{2004APS..APRS10001G,2004AIPC..727..637G,2005SSRv..120...95R}) from 2015 September 09 to 2015 October 06.  We take the {\it Swift} early {\it UBV} magnitudes from the {\it Swift} Optical/Ultraviolet Supernova Archive (SOUSA; \citealp{2014Ap&SS.354...89B}) and tabulate them in Table \ref{tab:observation_log_2015ap}.

The SNe instrumental magnitudes were calibrated with respect to the local standards in the SN field. Three Landolt standard fields PG 2331, PG 0918 and PG 0231 were observed along with the SNe fields with HFOSC/HCT to generate a sequence of local standards.  The observations were carried out at different altitudes with airmass varying between 1.1$-$2.0 with a typical seeing of $\leq$ 2 arcsec in $V$-band.  The instrumental and catalogue magnitudes of Landolt field stars were fitted,  using the least square regression technique, to estimate the zero points and colour term using a set of transformation equations \citep{1992JRASC..86...71S}. Site extinction values, taken from \cite{2008BASI...36..111S},  were used for the different filters.  A root-mean-squared (rms) scatter between transformed and standard magnitude of Landolt stars was found to be between 0.02 -- 0.05 mag in the {\it UBVRI} bands. Using the above zero points and colour term, we calibrated the magnitudes of 5 to 6 non-variable local standards in both the SNe fields (Table \ref{tab:optical_observations} and Table \ref{tab:optical_observations_2016P}). These secondary standards are used to calibrate the SNe instrumental magnitudes by applying nightly zero-points. The errors due to calibration and photometric measurement were added in quadrature to estimate the final error on the SNe magnitudes. Whereever required the {\it uri} magnitudes were converted to {\it URI} using the equations of \cite{2006A&A...460..339J}. The final photometry of SNe 2015ap and 2016P is given in Table \ref{tab:observation_log_2015ap} and Table \ref{tab:observation_log_2016P} respectively.

\begin{figure*}
	\begin{center}
		\hspace{-1.0cm}
		\includegraphics[width=0.5\textwidth]{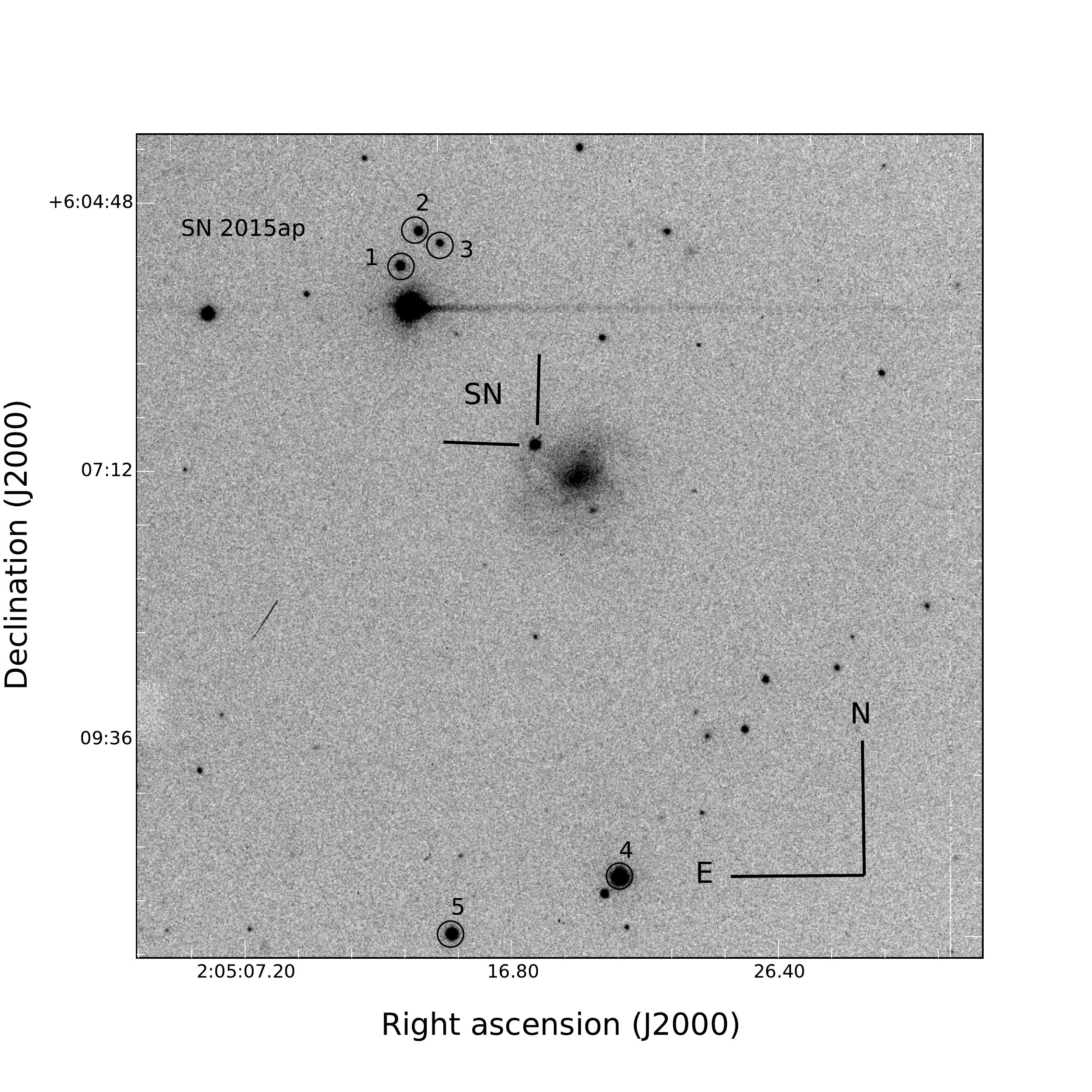}
		\includegraphics[width=0.5\textwidth]{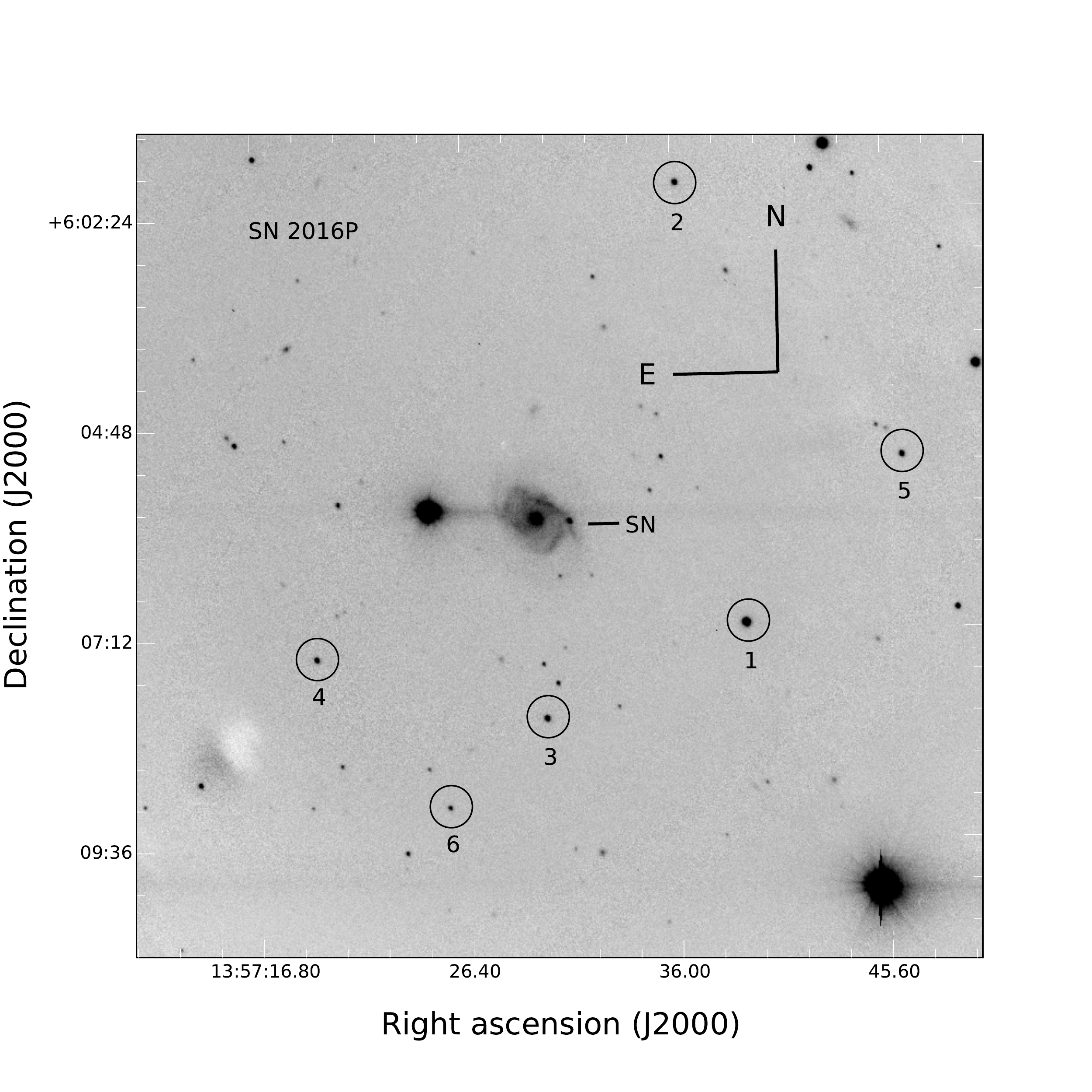} \\
	\end{center}
	\caption{SNe 2015ap and 2016P and the local standard stars in the field of IC 1776 and NGC 5374, $R$-band, 300 sec image obtained on 2015 September 26 and 2016 January 23 with the 2.0m HCT for SNe 2015ap and 2016P respectively.}
	\label{fig:calibimage_15ap}
\end{figure*}

\begin{table*}
\caption{Star ID and the magnitudes in {\it UBVRI} filters of the 5 secondary standards in the field of SN 2015ap}
\centering
\smallskip
\begin{tabular}{c c c c c c c c c}
\hline \hline
Star ID  &$\alpha$      &$\delta$         	& $U$   	    &  $B$              &  $V$          &  $R$          &  $I$                       \\
         & (h:m:s)      &($^{\circ}$ ' '')	& (mag)  	    & (mag)             & (mag)         & (mag)         & (mag)                     \\
\hline                           
1   &   02:05:08.19  & +06:04:32.52 &   17.91 $\pm$  0.09 &  17.843 $\pm$ 0.046  & 16.915 $\pm$ 0.056 & 16.712 $\pm$ 0.021 &  16.392 $\pm$ 0.442   \\     
2   &   02:05:07.31  & +06:04:29.29 &   17.47 $\pm$  0.04 &  16.887 $\pm$ 0.097  & 15.957 $\pm$ 0.081 & 15.604 $\pm$ 0.098 &  15.062 $\pm$ 0.101 \\     
3   &   02:05:06.96  & +06:04:49.94 &   17.12 $\pm$  0.03 &  16.495 $\pm$ 0.034  & 15.558 $\pm$ 0.089 & 15.406 $\pm$ 0.071 &  14.395 $\pm$ 0.098 \\     
4   &   02:05:19.97  & +06:09:48.75 &   15.01 $\pm$  0.02 &  13.685 $\pm$ 0.087  & 13.013 $\pm$ 0.091 & 12.712 $\pm$ 0.124 &  11.780 $\pm$ 0.142 \\ 
5   &   02:05:14.45  & +06:10:41.91 &   16.97 $\pm$  0.04 &  15.742 $\pm$ 0.088  & 14.597 $\pm$ 0.129 & 14.106 $\pm$ 0.112 &  3.055 $\pm$ 0.073 \\     
\hline                                   
\end{tabular}
\label{tab:optical_observations}      
\end{table*}

\begin{table*}
\caption{Star ID and the magnitudes in {\it BVRI} filters of the 6 secondary standards in the field of SN 2016P}
\centering
\smallskip
\begin{tabular}{c c c c c c c c}
\hline \hline
Star ID  &$\alpha$      &$\delta$         	   	    &  $B$              &  $V$          &  $R$          &  $I$                       \\
         & (h:m:s)      &($^{\circ}$ ' '')		    & (mag)             & (mag)         & (mag)         & (mag)                     \\
\hline                           
1  & 13:57:39.13 & +06:07:01.70  &	18.07 $\pm$  0.01   &   17.49 $\pm$ 0.02  &   17.12 $\pm$  0.01  &    16.79 $\pm$ 0.02   \\  
2  & 13:57:36.18 & +06:02:00.43  &	16.94 $\pm$  0.01   &   16.29 $\pm$ 0.01  &   15.87 $\pm$  0.01  &    15.51 $\pm$ 0.01     \\
3  & 13:57:30.03 & +06:08:05.81  &	18.83 $\pm$  0.02   &   17.33 $\pm$ 0.01  &   16.34 $\pm$  0.01  &    15.37 $\pm$ 0.01     \\
4  & 13:57:19.55 & +06:07:23.04  &	17.39 $\pm$  0.01   &   16.74 $\pm$ 0.01  &   16.32 $\pm$  0.01  &    15.94 $\pm$ 0.01     \\
5  & 13:57:23.44 & +06:09:35.95  &  	16.88 $\pm$  0.01   &   16.34 $\pm$ 0.01  &   15.99 $\pm$  0.01  &    15.67 $\pm$ 0.01     \\
6  & 13:57:48.83 & +06:06:52.69	 &	17.31 $\pm$  0.01   &   16.39 $\pm$ 0.01  &   15.83 $\pm$  0.01  &    15.36 $\pm$ 0.01      \\ 
\hline                                   
\end{tabular}
\label{tab:optical_observations_2016P}      
\end{table*}

\begin{center}
\begin{table*}
\footnotesize\addtolength{\tabcolsep}{-1pt}
\caption{Photometry of SN 2015ap} 
\begin{tabular}{c c c c c c c c c}
\hline \hline
 MJD          & $U$                 & $B$                    & $V$                 & $R$                    &  $I$       & Telescope\\    
              & (mag)               & (mag)                  & (mag)               & (mag)                  & (mag)     &            \\
\hline
   57282.47  &	15.04 $\pm$ 0.07   & 15.57 $\pm$ 0.06        & 15.29 $\pm$ 0.09    & ---                    & ---           & HCT \\
   57283.84  &  15.17 $\pm$ 0.06   & 15.63 $\pm$ 0.05        & 15.23 $\pm$ 0.01    & 15.14 $\pm$ 0.01       & 14.87 $\pm$ 0.01 &	 HCT \\
   57289.54  &	16.02 $\pm$ 0.08   & 15.94 $\pm$ 0.06        & 15.38 $\pm$ 0.06    & ---                    & ---          & ST \\
   57290.23  &  ---                & 16.43 $\pm$ 0.06        & 15.39 $\pm$ 0.06    & 15.15 $\pm$ 0.01 &   14.82 $\pm$ 0.01 &       HCT	\\
   57292.44 & ---                  & 16.51 $\pm$ 0.01        & 15.55 $\pm$ 0.07    & 15.23 $\pm$ 0.01 &   14.85 $\pm$ 0.03 &	      HCT	\\	
   57298.92  &	17.84 $\pm$ 0.19   & 17.28 $\pm$ 0.09        & 16.12 $\pm$ 0.08    & 15.66 $\pm$ 0.01 &   15.24 $\pm$ 0.01 &       ST	\\
   57299.83  &  ---                & 17.30 $\pm$ 0.17        & 16.18 $\pm$ 0.01    & 15.72 $\pm$ 0.01 &   15.23 $\pm$ 0.01 &         ST 	\\
   57300.92 & ---                  & ---                    &  16.15 $\pm$ 0.01    & 15.92 $\pm$ 0.01 &   15.44 $\pm$ 0.01 &	        ST	\\
   57302.44  &	18.01 $\pm$ 0.23   & 17.61 $\pm$ 0.11 &  16.33 $\pm$ 0.03 &   15.92 $\pm$ 0.02 &  15.43 $\pm$ 0.02 &       ST	\\
   57317.86 & ---                 & ---                    &  16.89 $\pm$ 0.02 &   ---                  &   16.44 $\pm$ 0.01 &	        ST	\\	
    57325.75 & ---                 & ---                    &  17.18 $\pm$ 0.36 &   16.82 $\pm$ 0.01 &   16.63 $\pm$ 0.01 &	        ST	\\
    57330.70 & ---                 & ---                    &  ---                &  17.12 $\pm$ 0.01 &  16.84 $\pm$ 0.01 &         ST	\\
    57331.77 & ---                 & ---                    &  17.25 $\pm$ 0.05 &   17.16 $\pm$ 0.01 &   16.84 $\pm$ 0.01 &         ST	\\
    57332.78 & ---                 & ---                    &  17.33 $\pm$ 0.03 &   17.17 $\pm$ 0.02 &   16.85 $\pm$ 0.02 &         ST	\\
    57334.72 & ---                 & ---                    &  ---                & 17.19 $\pm$ 0.02 &   16.95 $\pm$ 0.01 &         ST	\\
    57335.86 & ---                 & ---                    &  17.39 $\pm$ 0.05 &   17.19 $\pm$ 0.01 &   16.93 $\pm$ 0.02 &         ST	\\
    57336.83 & ---                 & ---                    &  17.45 $\pm$ 0.04 &   17.19 $\pm$ 0.02 &   16.94 $\pm$ 0.02 &         ST	\\
   57341.83 & ---                  &  18.26 $\pm$ 0.01      &  ---                &   ---                  &   ---                &         DFOT	\\
    57342.87 & ---                 & ---                    &  17.66 $\pm$ 0.04 &   ---   &    16.96 $\pm$ 0.02 &	        DFOT	\\
   57345.81 & ---                 & ---                    &  17.72 $\pm$ 0.02   &   17.28 $\pm$ 0.01 &   17.07 $\pm$ 0.02 &         ST	\\
   57353.78 & ---                   & 18.29  $\pm$    0.02 &  17.494 $\pm$ 0.019 &   ---                  &   17.24 $\pm$ 0.08 &	        HCT	\\
   57357.83 & ---                 & 18.43  $\pm$    0.03 &  17.79 $\pm$ 0.02 &  17.53 $\pm$ 0.02 & --- &	        HCT	\\
   57361.80 & ---                 & 18.54  $\pm$ 0.08 &  17.79 $\pm$ 0.05 &   ---                  &  17.29 $\pm$ 0.01 &         ST	\\
   57366.95   &	19.04 $\pm$ 0.04   & 18.36 $\pm$ 0.07 &  17.91 $\pm$ 0.06 &  17.82 $\pm$ 0.06 &   17.39 $\pm$ 0.06 &	        Copernico	\\
   57375.35 & 19.42 $\pm$ 0.08    & 18.61 $\pm$ 0.06       & 17.99 $\pm$ 0.09    & 17.95 $\pm$ 0.09 &  17.39 $\pm$ 0.09   & DFOT \\
   57386.69 & ---                 & ---                    &  18.28 $\pm$ 0.03 &   18.19 $\pm$ 0.02 &  17.70 $\pm$ 0.04 &         ST	\\	
   57387.70 & ---                 & ---                    &  18.29 $\pm$ 0.16 &   18.19 $\pm$ 0.02 &   17.72 $\pm$	0.05 &         ST	\\
   57390.62 & ---                 & ---                    &  ---                & 18.26 $\pm$ 0.03 &   17.78 $\pm$ 0.06 &         ST	\\
   57391.68 & ---                 & ---                    &  18.39 $\pm$ 0.09 &   ---                  &   17.79 $\pm$	0.05 &         ST	\\
   57392.68 & ---                 & ---                    &  18.02 $\pm$ 0.03 &   18.24 $\pm$ 0.04 &   17.80 $\pm$ 0.05     &         ST	\\
   57402.68 & ---                 & 18.79  $\pm$    0.09 &  18.58 $\pm$ 0.13 &   ---                  &   17.95 $\pm$ 0.05 &	        DFOT	\\
   57403.68 & ---                 & 18.78  $\pm$    0.06 &  18.59 $\pm$ 0.14 &   18.31 $\pm$ 0.05 &   18.08 $\pm$ 0.05 &	        DFOT         \\ 
   57412.59 & ---                   & 18.88 $\pm$ 0.02 &  18.69 $\pm$ 0.03 &   18.50 $\pm$ 0.20    &   18.201 $\pm$ 0.07 &	        HCT	\\
   57415.60 & ---                 & ---                    &  ---                & 18.48 $\pm$ 0.04   &   18.28 $\pm$ 0.06 &         ST	\\
   57417.64 & ---                 & ---                    &  18.76 $\pm$ 0.13 &   18.49 $\pm$ 0.05 &   18.30 $\pm$ 0.06 &         ST	\\
   57429.59 & ---                 & ---                    &  18.79 $\pm$ 0.09  & 18.59 $\pm$ 0.05 &   18.55 $\pm$ 0.01 &	        ST	\\
   57628.04   & 21.93 $\pm$ 0.09    & 21.50 $\pm$ 0.08  &  21.09 $\pm$ 0.08 &	  20.95 $\pm$ 0.08 & 20.86  $\pm$ 0.08 &	        Copernico	\\
\hline          		                    
\end{tabular}
\label{tab:observation_log_2015ap}     
\newline
\end{table*}
\end{center}

\begin{center}
\begin{table*}
\caption{Photometry of SN 2016P}  		  
\begin{tabular}{c c c c c c c c}
\hline \hline
MJD                             & $B$             & $V$           & $R$          &  $I$         & Telescope\\    
                               & (mag)           & (mag)         & (mag)        & (mag)        &            \\
\hline
	57408.00  &	    17.71 $\pm$ 0.05   &  16.98 $\pm$ 0.05 &	16.90 $\pm$ 0.03      & 16.74 $\pm$ 0.02  &		ST \\ 
        57408.98  &         ---                &  ---              &    16.69 $\pm$ 0.01      & 16.72 $\pm$ 0.03                   &          ST \\
	57409.52  &	    17.73 $\pm$ 0.09   &  16.98	$\pm$ 0.08 &    ---                   & 16.64 $\pm$ 0.02  &		ST  \\ 
	57410.91  &	    17.47 $\pm$ 0.06   &  16.90	$\pm$ 0.01 &    16.63 $\pm$ 0.04      & 16.43 $\pm$ 0.01  &		HCT \\
        57419.89  &	    18.05 $\pm$ 0.08   &  16.89 $\pm$ 0.09  &   16.41 $\pm$ 0.08      & 16.19 $\pm$ 0.01  &		HCT \\
        57420.99  &         ---                &  ---               &   16.44 $\pm$ 0.01      & 16.19 $\pm$ 0.01  &          HCT \\
	57422.02  &	    ---	               &  17.02 $\pm$ 0.01 &    16.49 $\pm$ 0.02      & 16.24 $\pm$ 0.01  &          HCT \\
	57424.98  &	    ---	               &  17.21 $\pm$ 0.01   & ---   		     &	16.30 $\pm$ 0.01  &	HCT \\	
	57428.91  &	    ---	               &  17.57 $\pm$ 0.05 &   16.99 $\pm$ 0.01      & 16.61 $\pm$ 0.03  &          ST  \\
	57432.00  &	    ---	               &  17.80	$\pm$ 0.07 &    ---                  & ---                 &          ST  \\
        57432.99  &         ---                &  17.83	$\pm$ 0.09 &    17.21 $\pm$ 0.08     & 16.74 $\pm$ 0.06   &          ST  \\
	57433.89  &	   19.02 $\pm$ 0.01    &  17.89 $\pm$ 0.09 &   17.35 $\pm$ 0.01      & 16.74 $\pm$ 0.01  &		HCT \\
	57438.98  &	    ---	               &  18.17 $\pm$ 0.01 &    17.73 $\pm$ 0.01  &  17.32 $\pm$ 0.03 &          HCT \\
	57448.79  &	    19.43 $\pm$ 0.01   &  18.31 $\pm$ 0.08 &    17.88 $\pm$ 0.01  &  17.81 $\pm$ 0.01  &		HCT \\
   	57451.50  &  	    19.45 $\pm$ 0.01   &  18.37 $\pm$ 0.07 &    17.99 $\pm$ 0.01  &  17.89 $\pm$ 0.01  &		HCT \\	
	57454.49  &	    ---	               &  18.46 $\pm$ 0.09 &    18.05 $\pm$ 0.01     & 17.81 $\pm$ 0.01  &		DFOT \\
	57454.74  &	    ---	               &  18.49 $\pm$ 0.09 &    18.07 $\pm$ 0.01     & 17.86 $\pm$ 0.01  &		DFOT \\
	57460.52    &	    ---	               &  18.74 $\pm$ 0.09 &    18.23 $\pm$ 0.01     & 18.04 $\pm$ 0.02  &          HCT \\
   	57461.80    &        ----              &  18.78 $\pm$ 0.04 &    18.26 $\pm$ 0.01     & 18.07 $\pm$ 0.01  &          HCT \\      
	57462.84    &	  19.58 $\pm$ 0.02     &  18.81 $\pm$ 0.02  &   18.26 $\pm$ 0.01     & 18.11 $\pm$ 0.08  &		ST  \\
	57465.90  &	    ---                &  ---                 & 18.30 $\pm$ 0.01     & 18.23 $\pm$ 0.02  &		ST  \\
	57478.32  &	    ---                &  19.00 $\pm$ 0.01 &    18.50 $\pm$ 0.01     & 18.51 $\pm$ 0.03   &          HCT \\
   	57485.54  &       ---	               &  19.26 $\pm$ 0.01 &    18.67 $\pm$ 0.01     & ---                 &      	HCT \\
   	57486.00  &	    ---                &  ---                 & ---   & 18.90 $\pm$ 0.02  & 		DFOT \\
        57489.50   &   20.12 $\pm$ 0.07 &  19.31 $\pm$ 0.07 &    18.93 $\pm$ 0.01  & 18.90  $\pm$ 0.01  &		DFOT \\
   	57491.41  &   20.18 $\pm$ 0.06 &  19.34 $\pm$ 0.01 &    18.99 $\pm$ 0.01  & ---  &		HCT \\
	57510.77  &	    ---                &  ---                 &    19.04 $\pm$ 0.01  & 19.16 $\pm$ 0.23  &		ST  \\
   	57542.67  &        ---                &  19.86 $\pm$ 0.08 &    19.41 $\pm$ 0.09  & 19.59 $\pm$ 0.01  &		HCT \\
\hline          		                    
\end{tabular}
\label{tab:observation_log_2016P}     
\end{table*}
\end{center} 		

\begin{table*}
\caption {Log of spectroscopic observations of SN 2015ap. The phase is measured with respect to $V$-maximum (MJD=57284.86$\pm$0.5).\label{tab:2015ap_spec_obs}}
\begin{center}
\begin{tabular}{c c c c c c}
\hline \hline
Date	   &     MJD        &    Phase  &    Telescope  &     Instrument  &		Range \\
           &                &           &               &                 &             \AA   \\
\hline    
2015/09/11  &	57276.06  &   -8.80   &  1.82m Copernico  &    AFOSC+gr04    	    &   3360-7740            \\
2015/09/18  &	57283.75  &   -1.11    &  2.0m HCT    &    HFOSC+gr07+gr08    &   3800-6840,5800-8350   \\
2015/09/24  &	57289.90  &    5.04    &  2.0m HCT    &    HFOSC+gr07         &   3800-6840             \\
2015/09/26  &	57291.94  &    7.08    &  2.0m HCT    &    HFOSC+gr07+gr08    &   3800-6840,5800-8350   \\
2015/10/08  &	57310.79  &    25.93   &  2.0m HCT    &    HFOSC+gr07+gr08    &   3800-6840,5800-8350 \\
2015/10/19  &	57314.67  &    29.81   &  2.0m HCT    &    HFOSC+gr07+gr08    &   3800-6840,5800-8350  \\
2015/10/28  &	57323.74  &    38.88   &  2.0m HCT    &    HFOSC+gr07         &   3800-6840            \\
2015/11/09  &	57335.64  &    50.78   &  2.0m HCT    &    HFOSC+gr07+gr08    &   3800-6840,5800-8350  \\
2015/11/27  &	57353.64  &    68.77   &  2.0m HCT    &    HFOSC+gr07+gr08    &   3800-6840,5800-8350 \\
2015/12/17  &	57373.86  &    89.00   &  1.82m Copernico  &    AFOSC+VPH06/07     &   3200-10000             \\
2015/12/20  &	57376.59  &    91.72   &  2.0m HCT    &    HFOSC+gr07+gr08    &   3800-6840,5800-8350     \\
2015/12/23  &	57379.57  &    94.71	 &  2.0m HCT    &    HFOSC+gr07         &   3800-6840               \\
2015/12/27  &	57383.59  &    98.73   &  2.0m HCT    &    HFOSC+gr07+gr08    &   3800-6840,5800-8350      \\
2016/01/22  &	57409.54  &   124.68   &  2.0m HCT    &    HFOSC+gr07+gr08    &   3800-6840,5800-8350  \\
\hline
\end{tabular}
\end{center}
\end{table*}          

\begin{table*}
\caption {Log of spectroscopic observation of SN 2016P. The phase is measured with respect to $V$-maximum (MJD=57416.75 $\pm$ 0.50).\label{tab:2016P_spec_obs}}
\begin{center}
\begin{tabular}{c c c c c c}
\hline \hline
Date	   &     MJD         &    Phase  &    Telescope  &     Instrument  &		Range \\
           &                &           &               &                 &             \AA   \\
\hline    
2016/01/23  &	57410.91  &   -5.84      &  2.0m HCT    &    HFOSC+gr07+gr08    &   3800-6840,5800-8350            \\
2016/02/01  &	57419.92  &   3.17      &  2.0m HCT    &    HFOSC+gr07+gr08    &   3800-6840,5800-8350   \\
2016/02/03  &	57421.90  &   5.15     &  2.0m HCT    &    HFOSC+gr07+gr08    &   3800-6840,5800-8350             \\
2016/02/05  &	57423.98  &   7.23     &  2.0m HCT    &    HFOSC+gr07+gr08    &   3800-6840,5800-8350   \\
2016/02/06  &	57424.91  &   8.16     &  2.0m HCT    &    HFOSC+gr07+gr08    &   3800-6840,5800-8350 \\
2016/02/20  &	57438.93  &   22.18     &  2.0m HCT    &    HFOSC+gr07+gr08    &   3800-6840,5800-8350  \\
\hline
\end{tabular}
\end{center}
\end{table*}



\bibliographystyle{mnras}
\bibliography{ref}

\bsp	
\label{lastpage}
\end{document}